\documentclass[]{spie}  

\usepackage{amsmath,amsfonts,amssymb}
\usepackage{graphicx}
\usepackage{amsmath}
\usepackage{float}

\usepackage[colorlinks=true, allcolors=blue]{hyperref}

\title{New developments on the Ingot WFS laboratory testing}

\author[a,b,d]{Tania S. G. Machado}
\author[a,d]{Simone Di Filippo}
\author[a,d]{Kalyan K. R. Santhakumari}
\author[a,d]{Maria Bergomi}
\author[a,d]{Davide Greggio}
\author[c,d]{Elisa Portaluri}
\author[b]{Dheeraj Malik}
\author[g]{Cesar Nesme}
\author[a,d]{Carmelo Arcidiacono}
\author[a,d]{Alessandro Ballone}
\author[a,b,d]{Federico Battaini}
\author[a,d]{Valentina Viotto}
\author[a,d]{Roberto Ragazzoni}
\author[a,d]{Marco Dima}
\author[a,d]{Luca Marafatto}
\author[a,d]{Jacopo Farinato}
\author[a,d]{Demetrio Magrin}
\author[a]{Luigi Lessio}
\author[e,f,a,d,b]{Gabriele Umbriaco}

\affil[a]{INAF – Osservatorio Astronomico di Padova, Vicolo dell’Osservatorio 5, 35122, Padova, Italy}
\affil[b]{Universit\`a degli Studi di Padova – Dipartimento di Fisica e Astronomia, Vicolo dell’Osservatorio 3, 35122, Padova, Italy}
\affil[c]{INAF – Osservatorio Astronomico d’Abruzzo, Via Mentore Maggini, 64100 Teramo, Italy}
\affil[d]{ADONI, Laboratorio Nazionale di Ottica Adattiva, Italy}
\affil[e]{Dipartimento di Fisica e Astronomia ”Augusto Righi” - Alma Mater Studiorum Universit\`a di Bologna, Via Piero Gobetti 93/2, 40129, Bologna, Italy}
\affil[f]{INAF Osservatorio di Astrofisica e scienza dello Spazio, Via Gobetti 93/3, 40129, Bologna, Italy}
\affil[g]{Ecole Centrale M\'editerran\'ee Marseille, 38 Rue Fr\'ed\'eric Joliot Curie, 13013, Marseille, France}

\authorinfo{Further author information: (Send correspondence to T.S.G.M)\\T.S.G.M.: E-mail: tania.gomesmachado@inaf.it, Telephone:+390498293410 \\  S.D.F.: E-mail: simone.difilippo@inaf.it, Telephone: +390498293519}

\begin{document} 
\maketitle

\begin{abstract}
The Ingot WFS was designed to overcome some of the challenges present in classical wavefront sensors when they deal with sodium LGSs. This innovative sensor works by sensing the full 3D volume of the elongated LGS and is suitable for use in very large telescopes. A test bench has been assembled at the INAF - Osservatorio Astronomico di Padova laboratories to test and characterize the functioning of the Ingot WFS. In this work, we summarize the main results of the tests performed on a new search algorithm. Then, we move towards a more accurate simulation of the sodium LGS by replicating real time-varying sodium layer profiles. The study of their impact on the ingot pupil signals is described in this work. 
\end{abstract}


\section{INTRODUCTION}
\label{sec:intro}  
\subsection{The Ingot Wavefront Sensor for LGS AO}

As the construction of Extremely Large Telescopes (ELTs) progresses, refining Adaptive Optics (AO) systems to maximize the benefits of Laser Guide Stars (LGS) is crucial. These telescopes will incorporate sodium LGS to enhance sky coverage for AO systems. However, the traditional wavefront sensors (WFSs) that are currently being planned to be used struggle with the elongated nature of LGS spots, presenting challenges yet to be solved.

A new class of wavefront sensors that can be deployed in a 3D manner and matches the geometry of the LGS was proposed by Ragazzoni et al. (2017)\cite{Ragazzoni2017}. These sensors can sense the cylindrical volume of LGS at finite distances, adapting to variable spatial light distributions and differential elongation in the pupil plane. This class of WFSs, named Ingot Wavefront Sensors (I-WFS), has various geometries based on different possible ways of splitting the beacon (from three to six pupils) \cite{Ragazzoni2018,Ragazzoni2024}.

In the current proceeding we use an I-WFS design with three pupils, which is the best compromise between loss of light and gain in sensitivity of wavefront derivative measurements \cite{Ragazzoni2019}.  In fact, this design copes with the continuously varying apparent width of the sodium layer and the resulting change of optimal length of the I-WFS portion, allowing the sharper end to sense the derivative of the wavefront orthogonal to the elongation.

\subsection{Sodium layer characteristics and variability}

Atomic sodium from meteoric disintegration forms a layer at altitudes between 89 km and 92 km. Astronomical observatories use lasers at $\lambda$ = 589.2 nm to excite this sodium, creating LGS. The detected sodium photon flux varies with the sodium column density, which fluctuates on seasonal, diurnal, and even 10-minute scales. Studies by Gardner et al. \cite{Gardner},\cite{gardner1990} revealed that the sodium layer's centroid height averages 92 km above mean sea level, with an average thickness of about 10 km FWHM. Significant variations in abundance (over 200\%), centroid height (2 km), and RMS width (1 km) have been observed within hours.

These variations cause defocus issues in WFSs designed for Natural Guide Stars (NGS). This work examines the impact of these variations on the three pupil I-WFS.

\subsection{Status of the project }

The project that aims to study the performance of the I-WFS is divided in several phases: elaboration of the concept \cite{Ragazzoni2019, Ragazzoni2024, Portaluri2024}; feasibility and sensitivity studies for the three pupils~\cite{Portaluri2022, Portaluri2023}; performance studies carried out by Portaluri et al. (2019) \cite{Portaluri2019} use numerical simulation tools\cite{Viotto2018}, which make use of a hybrid approach between a pure ray tracing technique and a Fourier analysis\cite{Viotto2019}; simulations of closed-loop procedures on an optical bench have also been carried out with a simplified model of the I-WFS \cite{Arcidiacono2020} and finally, laboratory tests and procedures will be further described in this work.

\section{METHODS}
\subsection{The test bench of the I-WFS}

\begin{figure}[H]
\begin{center}
\includegraphics[width=0.9\linewidth]{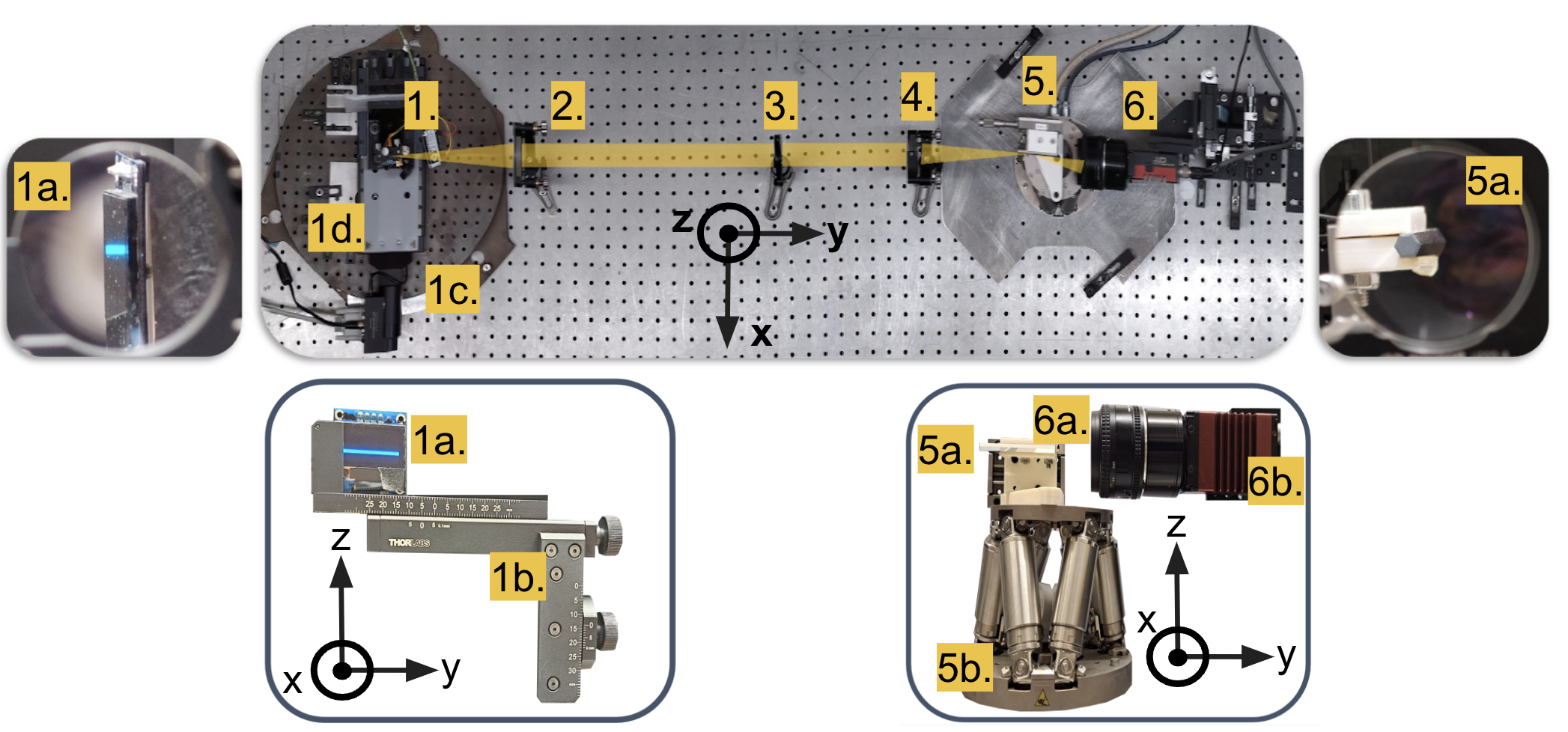}
\caption{Top view of the optical bench setup used to test the I-WFS. Detailed images of the OLED Screen (seen face on and from the pupil) and of the I-WFS (seen face on and from the pupil) can be seen on the left and right images, respectively. The legend of the images is as follows: 1. OLED Screen and support; 1a. OLED Screen with LGS; 1b. Vertical and Horizontal Linear Stage Thorlabs; 1c. Rotating platform; 1d. Linear Stage; 2. Collimating Achromatic Doublet f=200 mm; 3. Diaphragm D=25 mm; 4. Focusing Achromatic Doublet f= 200 mm; 5. I-WFS and support; 5a. I-WFS hexagonal light pipe 5b. H-811 Hexapod Physik Instrumente; 6. Pupil re-imager f= 50mm optics (6a.) attached to Prosilica GT3300 Camera from Allied Vision (6b.).  In this illustration, the orientation of the X, Y and Z axis used throughout this work is depicted. }
\label{fig:banco}
\end{center}
\end{figure}

The laboratory testing of the I-WFS is currently taking place at the INAF-Osservatorio Astronomico di Padova facilities. The experimental setup features a test bench specifically designed to assess the performance of an optical layout mimicking the E-ELT system with an I-WFS \cite{DiFilippo2019}. More precisely, the bench aims to simulate the LGS source, its variability and imaging onto the I-WFS in a 1:1 magnification optical layout\cite{DiFilippo2022}, generating an image of three pupils on a dedicated camera with accurate sampling. The test bench, illustrated on Figure \ref{fig:banco}, comprises readily available components now further described:

\begin{enumerate}
\item An OLED screen, programmed with an Arduino, replicates the LGS (1a.); 
\item The OLED screen is mounted on a Thorlabs linear stage that can be moved vertically (Z Axis) and horizontally along the optical axis (Y axis) (1b.) with $0.05$ mm precision and another linear stage (1d.) that allows movement perpendicular to the optical axis (X axis);
\item The OLED screen and linear stages are rested on a rotating platform that allows rotations with respect to the optical axis (1c.);
\item A collimating achromatic doublet with a focal length of $f = 200$ mm for the incoming LGS light (2.) and another identical one for refocusing the light onto the I-WFS prism (4.);
\item A diaphragm acting as the aperture stop/pupil with a clear aperture of $D=25$ mm (3.). Positioned for image space telecentricity, it is at the focus of the camera doublet;
\item A pupil re-imager optics employing a wide-aperture $f = 50$ mm photographic objective (6a.);
\item The I-WFS, built from a hexagonal light pipe, often used for beam homogenization (5a.). The external faces were aluminized to obtain a reflective roof with a 120 apex angle, matching the ELT requirement for proper pupil separation;
\item The camera, a Prosilica GT3300 from Allied Vision (6b.), equipped with an 8-megapixel CCD sensor featuring 5.5$\mu$m pixels. To sample the pupil diameter, a 4$\times$4 binning mode with 130 pixels is utilized;
\item The I-WFS prism is mounted on a H-811 Hexapod from \textit{Physik Instrumente} (5b.), enabling precise movements in all six degrees of freedom for the I-WFS;
\item In order to mitigate stray light effects, we added a black cover over the test bench.

\end{enumerate}

\vspace{-0.5cm}


\subsection{OLED Technology for LGS simulation}

OLED technology has been chosen to simulate the LGS after attempts with other types of displays, such as iPad screens and computer monitors. The primary advantage of OLED displays is their minimal background noise emission from black pixels. The second advantage is the ease with which the displays can be manipulated in an automatic way. However, a disadvantage is their tendency to wear out if left on for prolonged periods. Each screen is composed by a matrix of pixels that can be independently controlled via Arduino and integrated into Python procedures. To simulate the LGS, the screen is viewed edge-on and inclined relative to the pupil (as mentioned above), mimicking an LGS formed by a Laser Launcher Telescope (LLT) on the side of the main mirror, such as in the E-ELT. The left image labeled 5a. on Figure \ref{fig:banco} shows the view of the screen as seen from the pupil, where its elongated nature is already visible. The pixels along the Y direction of the screen, and hence at different distances from the pupil of the system, represent different altitudes in the sodium layer with respect to the pupil of the telescope.


Initially, an \textbf{SSD1306 Monochrome OLED display} (named \textbf{old screen}, from hereafter) was used to simulate the source. A detailed description of how a source was characterized and of the tests performed with it can be found in Ref.~\citenum{machado2023}. The main limitation of this model of OLED is that it can only be used to create sources with a uniform brightness across all pixels.  Since a real LGS originates from a layer with varying Na concentrations, a realistic simulation of a LGS profile should involve different brightness along the length of the LGS on the screen. This variability is both spatial, changing at different altitudes along the sodium layer, and temporal, as the LGS intensity changes over time. To understand the impact of these fluctuations on the I-WFS procedures, we have adopted a \textbf{new OLED screen model}, the \textbf{SSD1327}, which supports a \textbf{16-level greyscale} display (named \textbf{new screen}, from hereafter). This allows to set different brightness levels on each pixel of the screen on a scale from 0 to 15. The setting of greyscale automatically performs a correction on the OLED panel display so that the perception of the brightness scale shall match the user input value. As a consequence, the brightness of each pixel is linear with the greyscale value.

\begin{figure}[H]
\begin{center}
\includegraphics[width=0.8\linewidth]{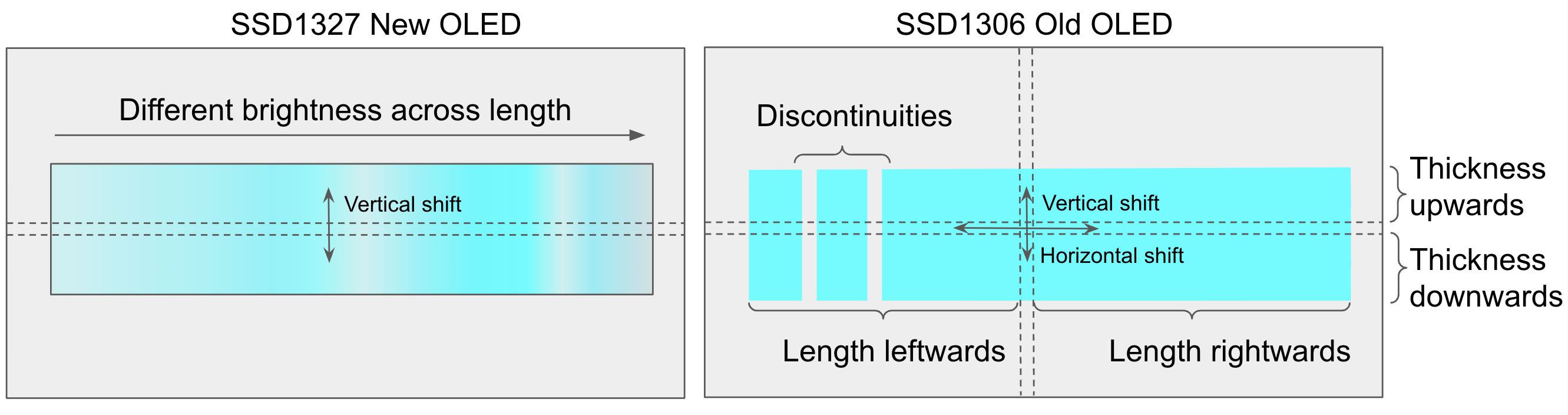}
\caption{A schematic illustration of the two different OLED models is shown and the process of creating an LGS is exemplified, along with the possible changes that can be applied to the sources. (Left) SSD1327 16-level greyscale  OLED screen with LGS with different brightness along its length and possibility to change the vertical position on the screen. (Right) SSD1306 Monochrome OLED screen with LGS characterized by seven parameters: vertical shift, horizontal shift, upward thickness, downward thickness, leftward length, rightward length, number of discontinuities.}
\label{fig:comparing_screens}
\end{center}
\end{figure}

In Figure \ref{fig:comparing_screens}, a schematic illustration of the two different OLED models is shown. On the left illustration, a source on the new screen is drawn with a differential brightness across its length, \textit{i.e.} over the pixels along the Y direction. There is still the possibility of shifting this source vertically on the screen, by illuminating different rows. On the right illustration, a source on the old screen is drawn with all the parameters used to define it, as further explained in Ref.~\citenum{machado2023}. 

\vspace{-0.5cm}
It will be important for the remainder of this work to define a standard source, usually used during calibration procedures, which represents the most basic configuration. The standard source is designed to simulate a LGS with 1.5” of thickness and 6.05” of length on sky. On the new OLED screen model, in order to keep the same length of LGS, and so simulate a sodium layer with the same thickness, the resulting area of the standard source is $20.9 \textrm{mm}^ 2$ whereas it was $14.4\textrm{mm}^ 2$ with the old screen.

\subsection{The I-WFS alignment procedure}

Most of the methodologies incorporated in the data analysis have been perfected and enhanced up until this point. This encompasses the pupil analysis procedure, the absolute alignment procedure, and the calibration procedure. These methods can be found described in more detail in Ref.~\citenum{DiFilippo2022} and ~\citenum{2020SPIEKalyan}.


\begin{figure}[H]
\begin{center}
\begin{tabular}{cc}
\includegraphics[width=0.3\linewidth]{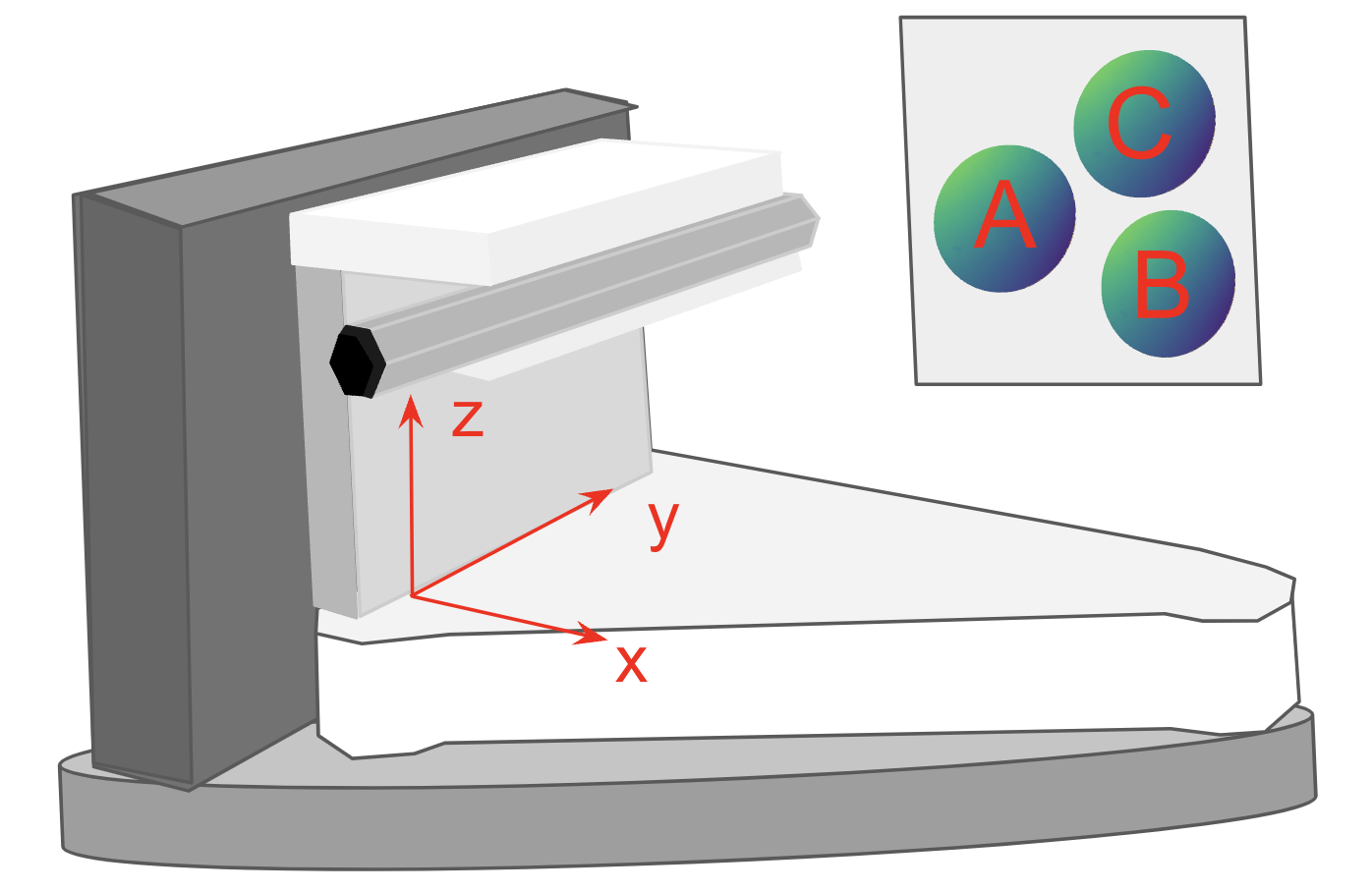} & \includegraphics[width=0.55\linewidth]{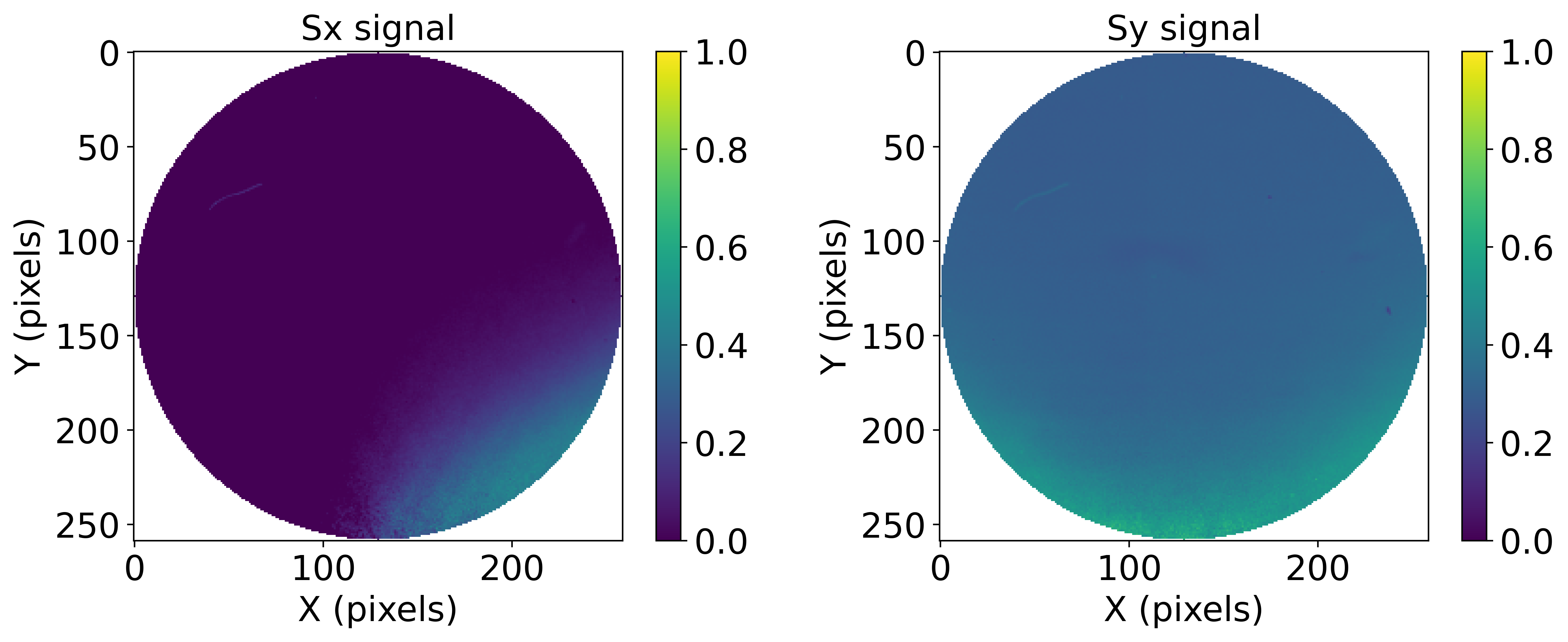}
\end{tabular}

\caption{(Left)The I-WFS hexagonal pipe holding structure, designed such that only two of the six faces intersect the incoming light. This holding structure can be moved by the hexapod in its six degrees of freedom and the orientation of the axis X,Y and Z is also shown here. The three pupils created by the I-WFS are illustrated, showing their relative orientation. (Right) Signals $S_x$ and $S_y$ are calculated from the pupils and are here displayed, for an aligned configuration.\label{fig:ingot_structure}  }

\end{center}
\end{figure}

\vspace{-0.5cm}

In Figure \ref{fig:ingot_structure} we show and identify the three pupils formed by the I-WFS and re-imaging optics on the camera. Pupil A is the transmitted pupil, while pupils B and C are the reflected pupils. Once the pupils are located and aligned, they are used to calculate the signals $S_x$ and $S_y$ pixel-wise~\cite{DiFilippo2022}. From the pupils data and the signals $S_{x}$ and $S_{y}$ a set of six observables was defined in order to establish the ideal aligned configuration of the system. Hence, the aligned position is identified by a target value corresponding to each of the six observables, as stated on Table \ref{tab:observables}.


The alignment procedure is based on a linear approximation approach similar to that commonly used in AO closed-loop theory. Interaction and Control matrices (IM and CM), that are typically used with deformable mirrors, are calculated in a dedicated calibration routine implemented in Python. The alignment procedure consists of an iterative method that starts with a misaligned ingot prism and executes incremental movements to converge to a better aligned position, using the previously obtained CM and IM. The alignment process continues running as long as the observables fall outside a predefined tolerance range. The alignment tolerance values defined for each observable are written on Table \ref{tab:observables}.

\begin{table}[ht]
\caption{The target values of each of the six observables that define the aligned position are presented. The tolerance of each observable that define the aligned range are also specified.} 
\label{tab:observables}
\begin{center}       
\begin{tabular}{|c|c|c|} 
\hline
\rule[-1ex]{0pt}{3.5ex} Observable & Target Values $\pm$ Tolerance  \\
\hline
\rule[-1ex]{0pt}{3.5ex}  Flux$(C-B)/\textrm{Flux}(A+B+C)$  &  $0\pm0.5$ \%        \\
\hline
\rule[-1ex]{0pt}{3.5ex}  		Flux$(A-B-C)/\textrm{Flux}(A+B+C)$   & $-33.33\pm0.4$\%      \\

\hline
\rule[-1ex]{0pt}{3.5ex}  Distance $A-B$    & $277\pm1$ px        \\

\hline
\rule[-1ex]{0pt}{3.5ex}  Distance $A-C$     & $277\pm1$ px        \\

\hline
\rule[-1ex]{0pt}{3.5ex}  Distance $Y_B-Y_C$  & $0\pm1$ px     \\

\hline
\rule[-1ex]{0pt}{3.5ex}   $\overline{S_{x1}}  - \overline{S_{x2}} $    & $0\pm0.2$ \%         \\
\hline 
\end{tabular}
\end{center}
\end{table}

\vspace{-0.5cm}
\subsection{Sodium layer real data}

Authentic sodium layer profiles were obtained at the 6-metre Large Zenith Telescope in Canada \cite{Pfrommer:10} using LIDAR technology, and very kindly made available to us by Angel Otarola from the TMT International Observatory. Using LIDAR technology (Laser Imaging Detection and Ranging), information regarding the sodium density at different altitudes can be obtained \cite{Gardner}. This method consists in measuring the time between the emission of laser pulses tuned to excite the sodium atoms and the detection of the backscattered signals, which will provide insight into the altitude and density of the impacted atoms. The data set comprises a total of 16 cases, obtained throughout the month of July 2010, with each case consisting of 30 profiles. These profiles are the result of 10-second time integrations, with an interval of 5 minutes between each measurement, and offer a vertical resolution of 126.5 metres. The data set encompasses 198 height levels, starting from the lowest altitude bin of 80 km. 

\begin{figure}[H]

\begin{tabular}{c} 
\hspace{-0.3cm}
   \includegraphics[width=\linewidth]{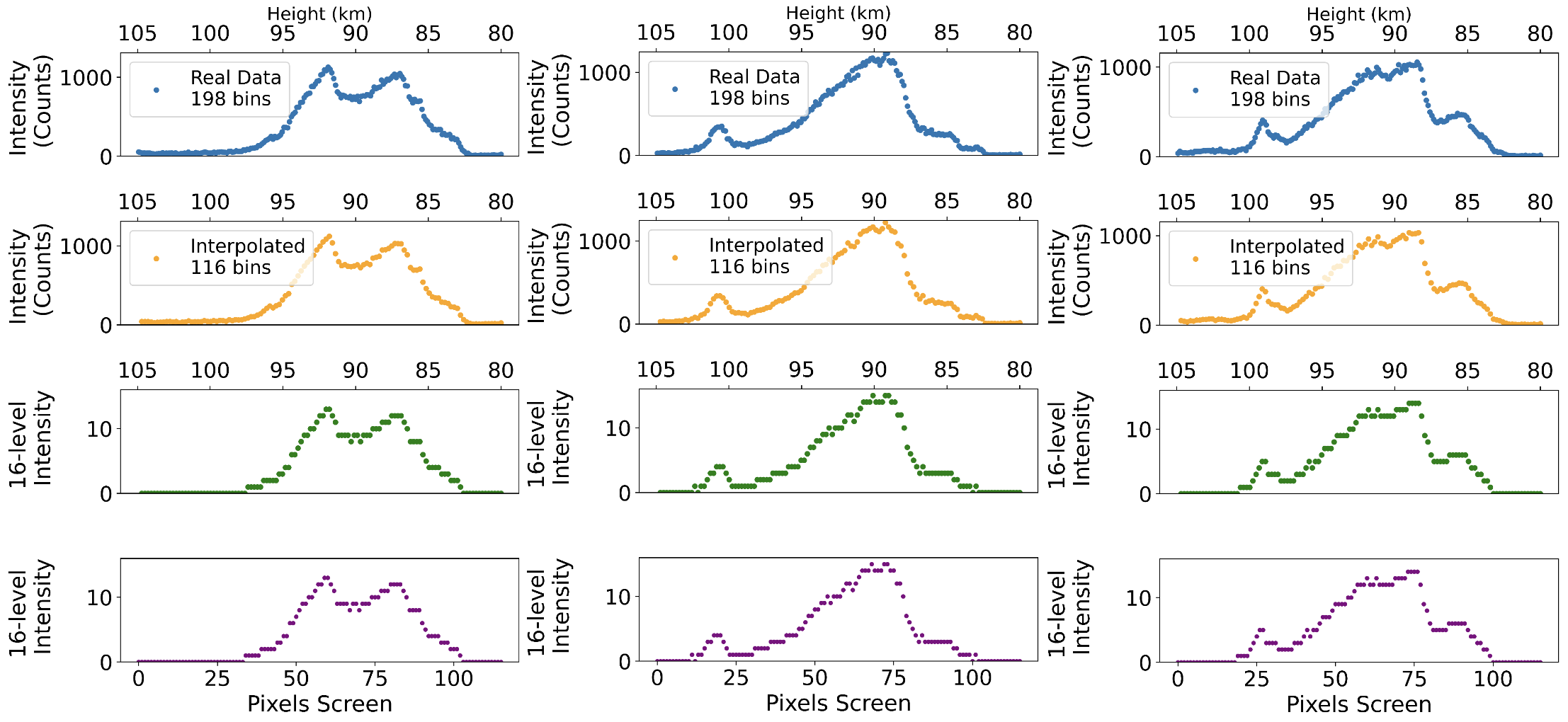} 
   \end{tabular}

   \caption[] 
   { \label{fig:lgs_layer_replicate}  Plots illustrating the process of replicating sodium layer profiles onto the LGS on the new OLED screen. (Left) Case 1 profile taken at the beginning of the night. (Center) Case 2 profile taken during the night, 2.5 hours after Case 1. (Right) Case 3 profile taken 2.5 hours after Case 2.}

\end{figure}

The data made available to us is of use to simulate realistic sodium layer profiles along the extension of the LGS on the OLED screen. Present on the top panels of Figure \ref{fig:lgs_layer_replicate} are three of the cases of this set of data, drawn from the same night but separated by 2.5 hours. From left to right, they will be referred to as Case 1, Case 2 and Case 3. The intensity of the sodium concentration is plotted as a function of heigh in km, for all the 198 height bins. In order to replicate these profiles on the OLED screen, the data needs to be re-binned down to only 116 height bins and  intensity of the profile must be translated into one of the 16 brightness levels (an integer from 0 to 15). The steps of this data treatment are illustrated on Figure \ref{fig:lgs_layer_replicate}, where the bottom panels show the data now ready to be passed on to the screen.


Below, in Figure \ref{fig:real_3_cases}, these three cases are displayed on the screen, respectively from left to right in crescent order.

\begin{figure}[H]
\begin{center}
\begin{tabular}{ccc} 
   \includegraphics[width=4.5cm]{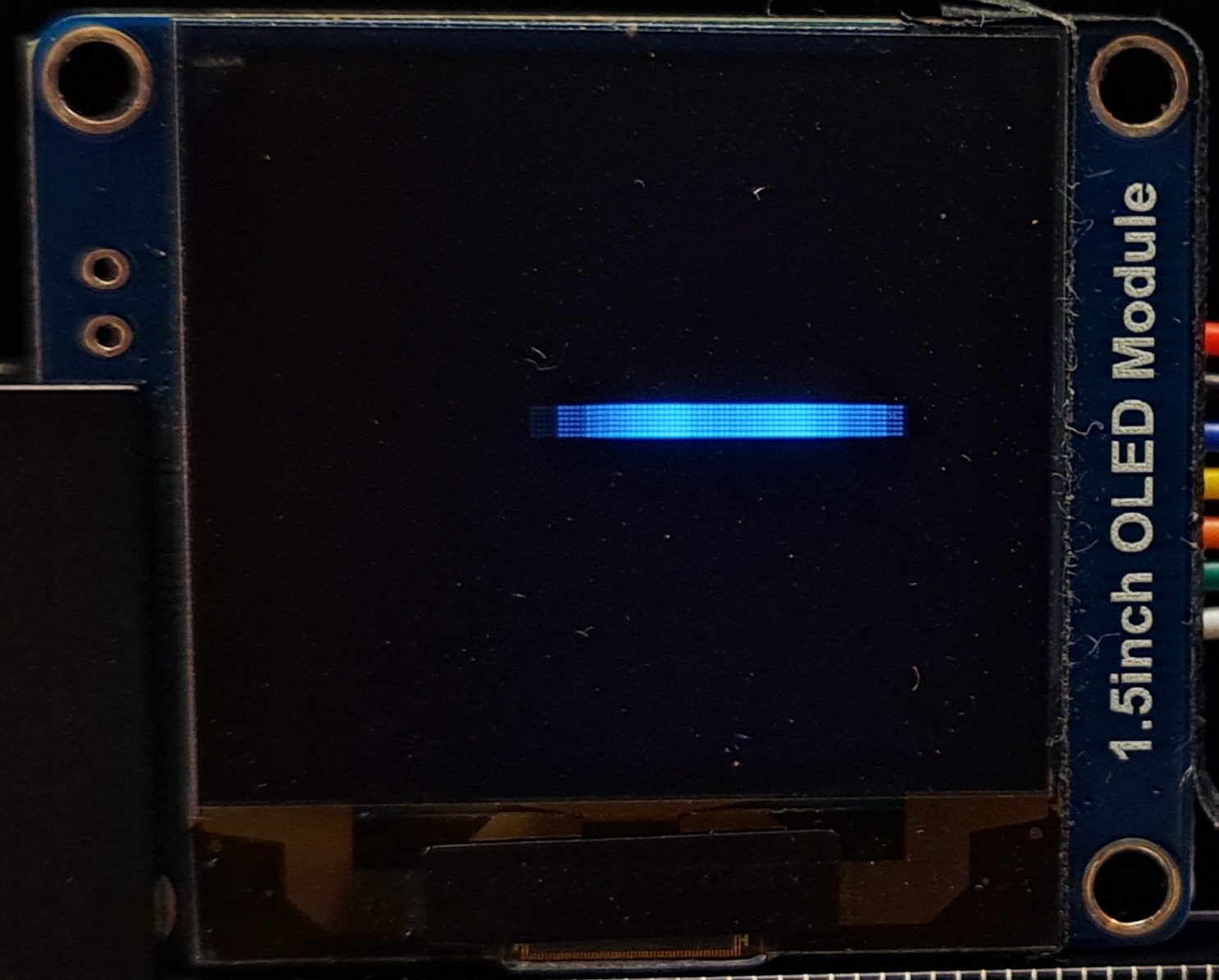} &
\includegraphics[width=4.5cm]{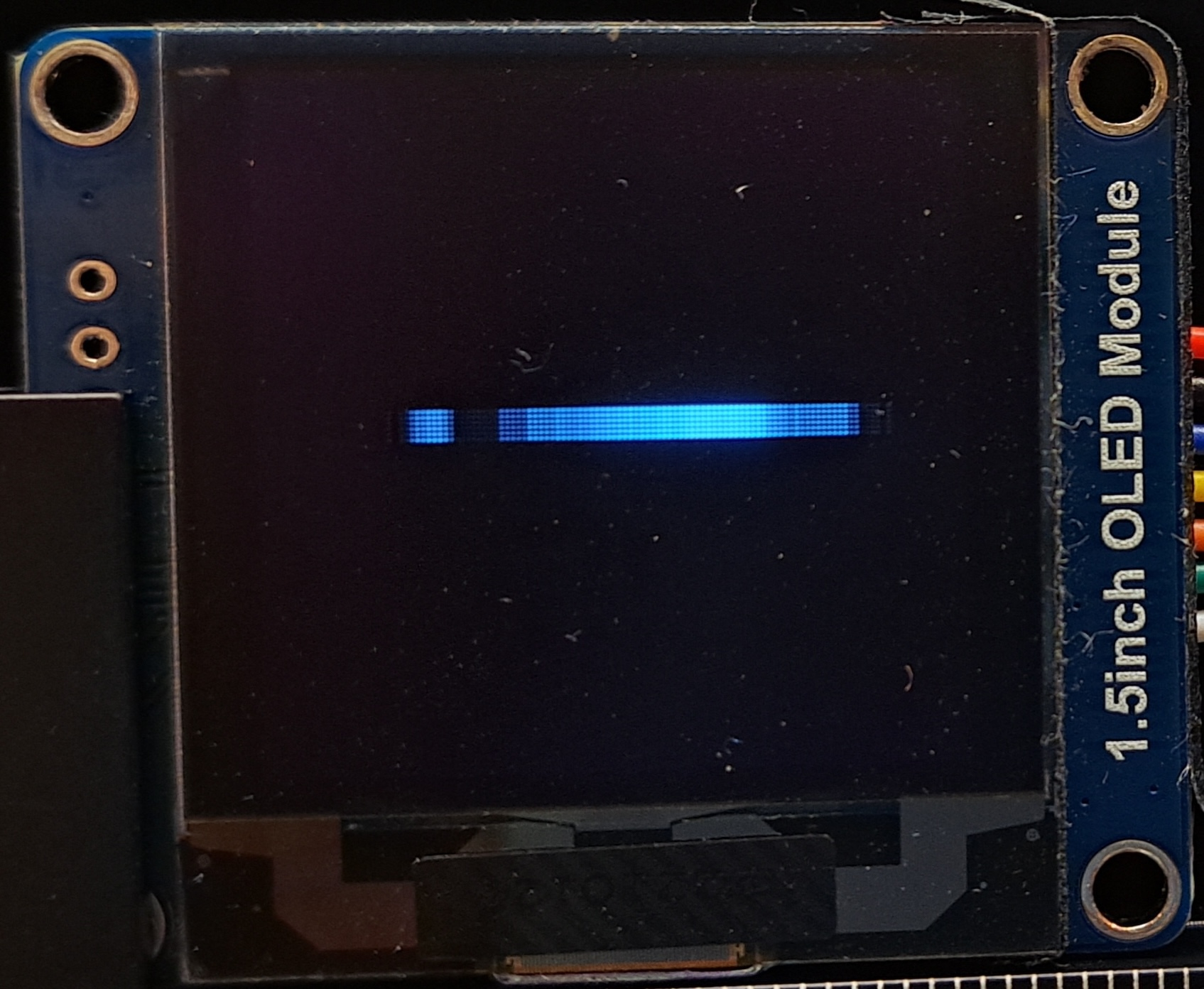} & \includegraphics[width=4.5cm]{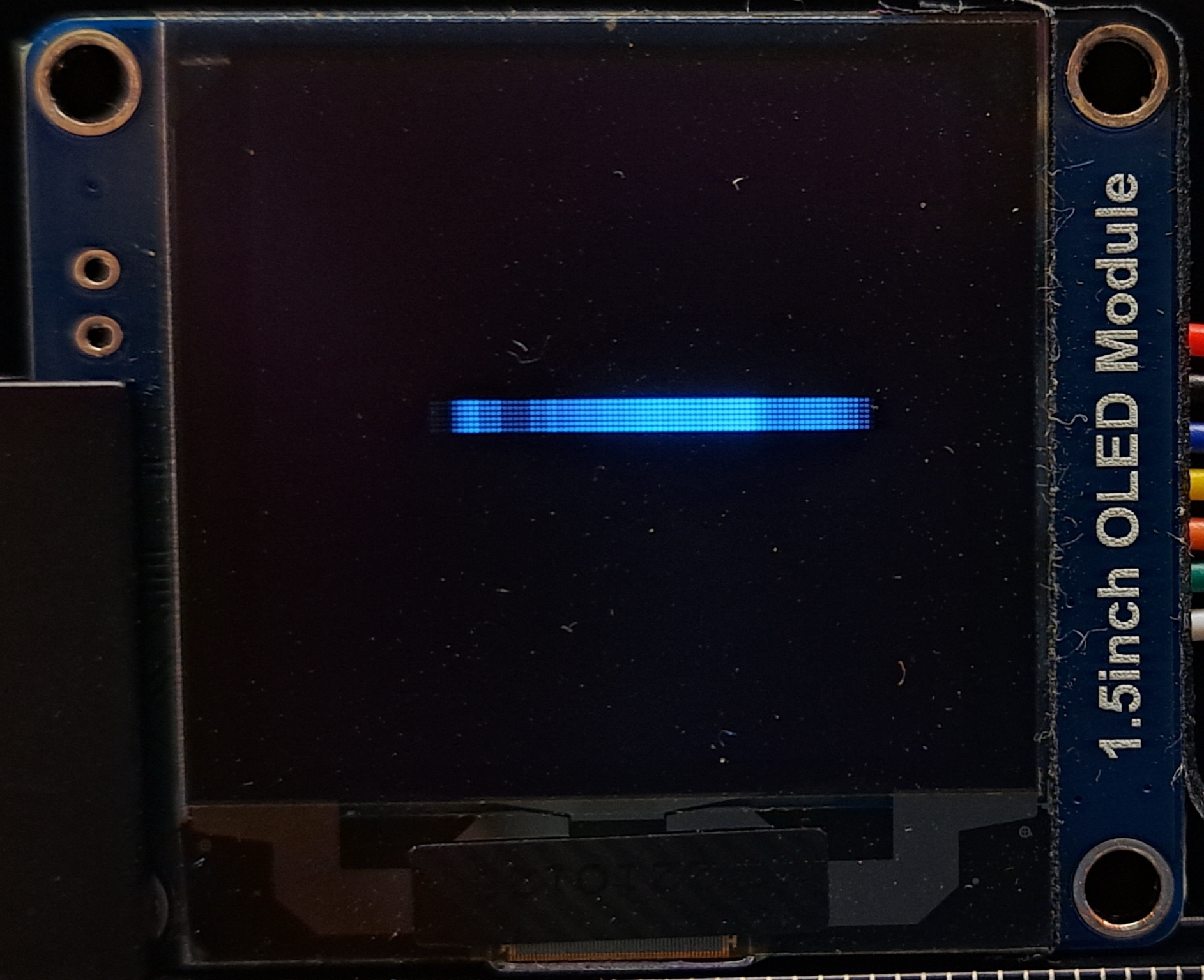}
   \end{tabular}

   \caption[] 
   { \label{fig:real_3_cases}  Images of the three cases shown above in Figure \ref{fig:lgs_layer_replicate} when displayed on the new OLED screen. (Left) Case 1 profile taken at the beginning of the night. (Center) Case 2 profile taken during the night, 2.5 hours after Case 1. (Right) Case 3 profile taken 2.5 hours after Case 2. }

\end{center}
\end{figure}

\vspace{-0.5cm}
\section{RESULTS}
\subsection{The search algorithm}

\begin{figure}[H]
\begin{center}
\begin{tabular}{cc} 
    \includegraphics[width=0.3\linewidth]{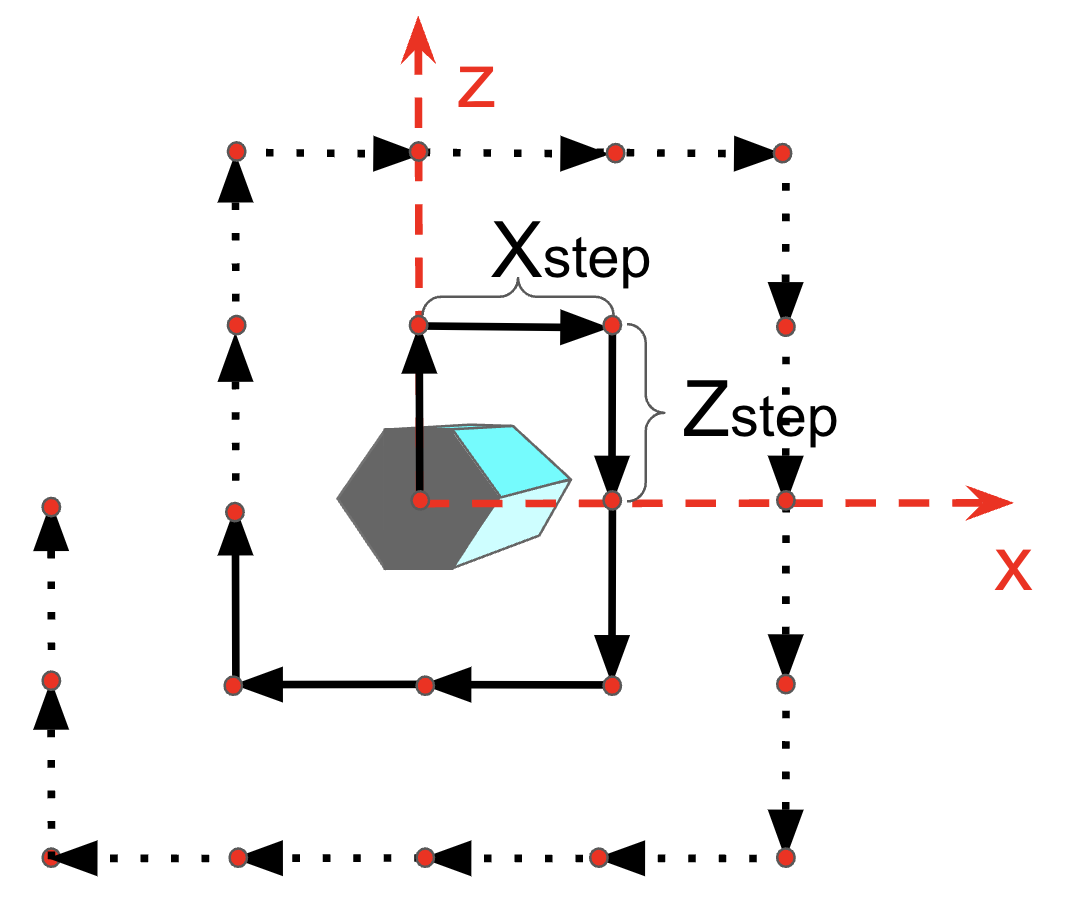} &  \includegraphics[width=0.6\linewidth]{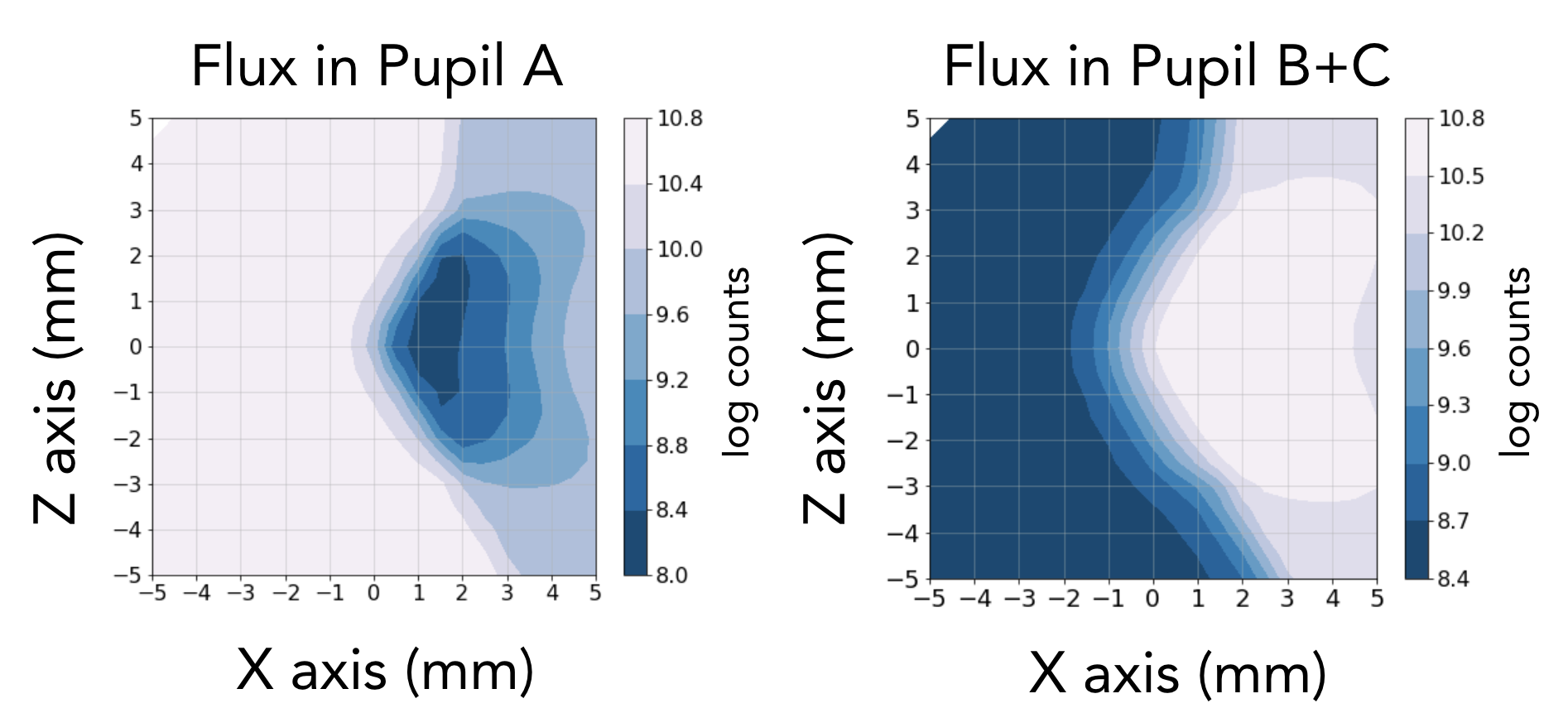} \\
   \end{tabular}

   \caption[] 
   { \label{fig:search_movement} (Left) The search algorithm scan follows a specific direction on the X-Z plane, taking steps of size $X_{step}$ and $Z_{step}$. (Right) Contour plots show the scan performed on the X-Z plane, with a resolution of 0.1 mm, revealing the logarithmic counts on pupil A and the logarithmic sum of counts on pupil B and C. Images adapted from Gomes Machado Master Thesis \protect\footnotemark.}

\end{center}
\end{figure}

The search algorithm is part of a broader initiative that aims to create a system capable of detecting light even in cases where the prism initially lacks any illumination. When the I-WFS experiences significant misalignment, it may even lead to a situation where only one or two pupils become illuminated. A critical consideration is whether the I-WFS can recover from these scenarios, as they could realistically occur during telescope operations. Beginning with an absence of illuminated pupils, the objective is to advance through a sequence: from 0 pupils to 1 pupil, then from 2 to 3 and ultimately to achieve a viable alignment setup. The search is done in a spiral movement of the I-WFS, as illustrated on Figure \ref{fig:search_movement}.

\footnotetext{\href{https://hdl.handle.net/20.500.12608/51828}{Link to thesis.}}
\vspace{-0.5cm}
The proposed strategy involves primarily restricting the initial movement to the X-Z plane, as it should be enough to intersect the light beam. Figure \ref{fig:search_movement} (right), depicts an initial scan of the entire X-Z plane within a region of -5 mm to 5 mm, centred at zero, that was used to understand which strategy to adopt. The values of the total counts in pupils A, B and C were recovered with a spatial resolution of 0.1 mm. From this analysis, we developed a search algorithm divided in two parts:
\begin{enumerate}
\item From 0 to 1 illuminated pupil (pupil A): when there is no light reaching the transmitted pupil A, the search algorithm will use the counts inside this pupil as target to maximise throughout its spiral search.
\item From 1 illuminated pupil to 3 illuminated pupils (pupils B and C): when there is no light reaching the reflected pupils B and C, the search algorithm will use the sum of counts inside these pupil as target to maximise throughout its spiral search.
\end{enumerate}

Currently we are limited to the amount of different scenarios for which we can test this search algorithm due to the limited range of motion of the hexapod stage. Positive results were obtained with the current tests nevertheless.

\subsection{Refining the alignment procedure}

\begin{figure}[H]
\begin{center}
\begin{tabular}{c} 
   
\includegraphics[width=\linewidth]{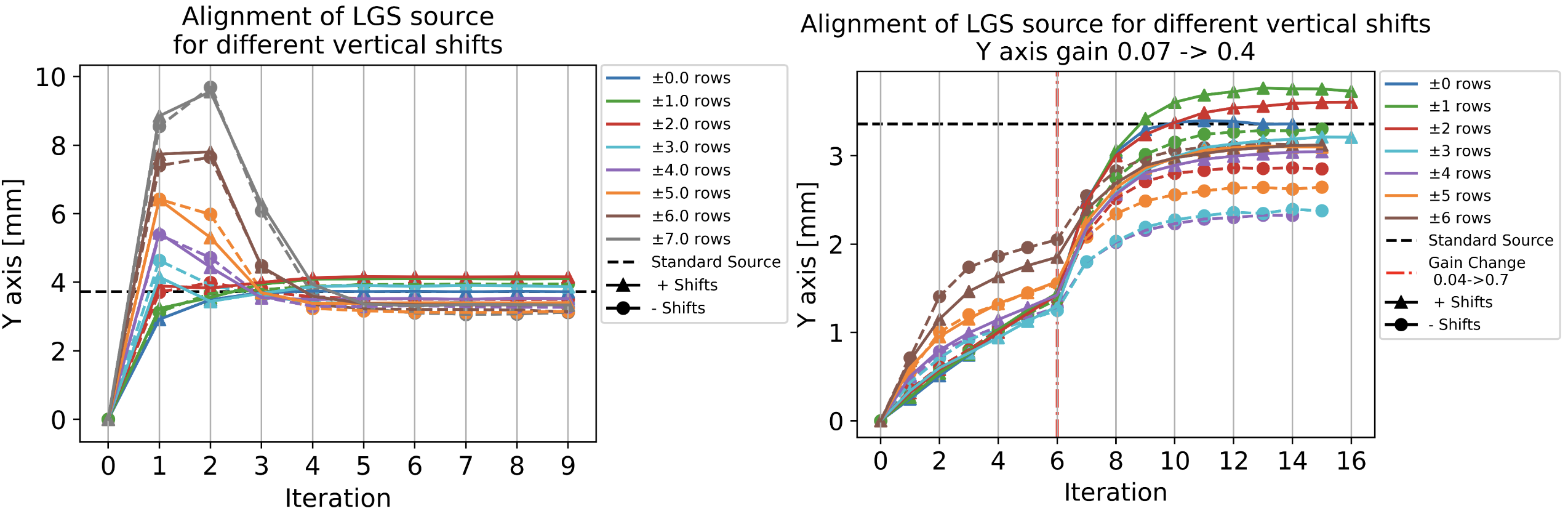}
   \end{tabular}

   \caption[example] 
   { \label{fig:yovershoot} 
(Left) Plot depicts the progression of the Y position of the hexapod/ingot during the alignment process when different sources are placed on the screen. The y-axis represents the position on the Y axis of the hexapod, in mm, while the x-axis displays the iteration when that position was achieved. Each colour represents a source with different values of vertical shift (VS) of rows on the OLED Screen: positive VS are shown with continuous lines and triangular scatter points, while negative VS are represented with dashed lines and circular scatter points. The final position of the standard source Y$_{standard} (VS = 0)$ is highlighted. The overshoot behaviour on this axis is visible, reaching values up to 10 mm before converging to a smaller value. (Right) Applying a different gain on the Y axis, the Y overshoot behaviour has been corrected. The same study as the adjacent plot was performed, fixing the gain to 0.07 during the first 5 iterations and changing to 0.4 for the remaining iterations. }

\end{center}
\end{figure}

Throughout the tests conducted over the past months, with the old OLED screen still in place, a behaviour termed "Y axis overshoot" consistently emerged during different alignment tests. This phenomenon, exemplified in Figure \ref{fig:yovershoot}, is characterized by $\Delta Y = 10$ mm  followed by an inversion in direction. This overshoot poses a problem for the current system since a movement of this order is very close from reaching the limit of the hexapod range of motion in this axis. This phenomenon stems from the ingot higher sensitivity on the Y axis whenever there is a flux change in the pupils. In fact, axis Y and X both affect the observable Flux$(A-B-C)$ but a movement of $0.1$mm on Y axis is approximately $0.37$ times less sensitive than a movement of $0.1$ mm on the X axis. Moreover, this hints at a degeneracy between the observables and the coordinates as the Y axis should be mostly related to horizontal movements rather than vertical movements of the source.

To address this behavior, we tested different gains applied to each degree of freedom. Initially, and up until now, the gain for each degree of freedom had been set to 0.7. This gain is used to calculate the positioning corrections of the I-WFS at each iteration to bring it closer to an aligned position. After several tests, we determined that the optimal gain combination to mitigate the Y axis overshoot without significantly increasing the number of iterations was a small gain for the first five iterations (0.07), followed by an increase to 0.4, for the Y axis only. The other degrees of freedom retained a fixed gain of 0.7 throughout all iterations. Figure \ref{fig:yovershoot} shows the improvement in alignment with this gain adjustment.

\vspace{-0.5cm}
Two other modifications were implemented: increasing the tolerance for the last observable from 0.1 to 0.2, and adding a stability condition to the alignment process. This stability condition requires three consecutive iterations to be within the aligned range tolerance, ensuring the system is stably aligned and not merely by chance. All these changes have resulted in an increase of the average number of iterations from 8 to 16.

\subsection{Comparison between OLED screens}

A full comparison of the two OLED models was performed before the setup was definitely changed to the newly acquired OLED Screen. To be able to compare the two screens simultaneously, all tests were performed with a so called "short source" (reduction of 50\% in length), instead of the usual standard source. A change in screen, source, and position of the screen requires attention regarding the choice of the reference position of the calibration procedure, which should be very close to an aligned position. Hence, after carefully manually selecting a new reference position for the calibrations of this short source with both screens, the following tests were condone:

\begin{enumerate}
\item \textbf{Alignment of the short source:} The system was able to reach an aligned position. 
\item \textbf{Study of the background emission originating from dark pixels of the screens:} 
The old OLED screen (connected to power but without any lit pixel) introduces an increase in counts on the dark images taken with the camera of 0.7\%, whereas the new OLED screen introduces an increase of 0.5\%.This increase is always considered to be relative to the absence of any screen on the optical path.

\item \textbf{Intensity of the emission spectra of the two screens:}
Using an \textit{Ocean Insight} spectrometer (model HDX-XR, spectral resolution $0.550$ nm), we measured the spectral emission of the two screens by placing an optical fiber in the optical path of the system. The resulting emission normalized to the maximum counts on each screen is shown Figure \ref{fig:spectra_compare}. The wavelength of maximum emission is highlighted: $476.587$ nm and  $461.261$ nm respectively for the old and new models. Furthermore, the new screen emission curve seems to be a composition of three broader peaks of emission located around the red, green and blue wavelengths. This is explained due to the fact that each OLED pixels is composed of an RGB matrix. Moreover, this emission also explains why visually the screen looks whiter than the old screen, which has a peak of emission mostly around the blue wavelength. Finally, on Figure \ref{fig:spectra_compare} we have also highlighted the wavelength corresponding to the LGS real wavelength due to the sodium backscattering. At that wavelength, the new screen emits five times more than the old screen. This could be of relevance in a future update of the bench where only the Na wavelength is selected for a more accurate simulation of the real LGS.

\begin{figure}[H]
\begin{center}
\includegraphics[width=14cm]{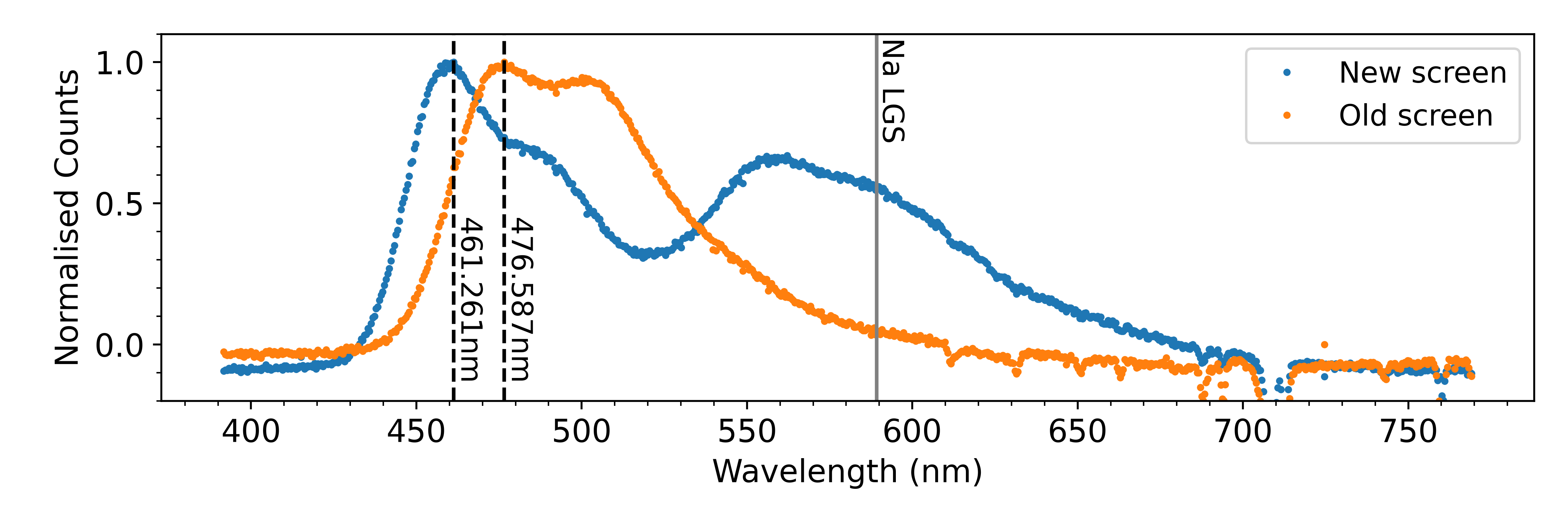}
\caption{Normalized emission as a function of wavelength for the old and new OLED screens, taken by inserting a fiber in the optical path of our test bench. The wavelength of maximum emission is highlighted: $476.587$ nm and  $461.261$ nm respectively for the old and new models. We have also highlighted the wavelength corresponding to the LGS real emitted wavelength ($\lambda = 589.2$ nm).}
\label{fig:spectra_compare}
\end{center}
\end{figure}

\end{enumerate}

\vspace{-0.5cm}
After deciding to definitely change the setup to fix the new OLED screen, we repeated some of the previous tests, now with the standard source, and proceeded to characterise other aspects of the screen.

\begin{enumerate}
\item \textbf{Flux and SNR of the aligned pupils:}
After successful calibration and alignment of the standard source on both screens, the flux on the aligned pupils was compared. In addition, the flux on the transmitted pupil obtained by removing the I-WFS from the optical path was also measured. These fluxes can be used to state that the new screen is creating an increase of factor 1.7 in the transmitted pupil flux and 1.85 on the aligned pupils, relative to the old screen. These values were obtained considering that the standard sources on both screens do not have the same emitting area. On the other hand, these fluxes also allowed us to conclude that inserting the I-WFS on the optical path is resulting in a loss of 28.5\% of light, either scattered or reflected away from the camera, for the case of the old screen and 22.12\% for the case of the new screen.

Then, the signal to noise ratio (SNR) of the images was computed and the average SNR calculated on each pupil is shown on Table \ref{tab:snr_standard}.  As noise contribution we considered the dark frames (which are always subtracted to every frame) obtained when no source is illuminated on the screen and we neglected the read out noise since the data obtained is far from a low brightness regime. From these values, we again confirm that the new version of OLED results on averaged SNR twice as big as the old version.

\begin{table}[ht]
\caption{Comparison of averaged SNR measured on the three pupils with the standard source on the old and new OLED screen. } 
\label{tab:snr_standard}
\begin{center}       
\begin{tabular}{|c|c|c|} 
\hline
\rule[-1ex]{0pt}{3.5ex}  SNR & Old OLED Screen & New OLED Screen  \\
\hline
\rule[-1ex]{0pt}{3.5ex}  Pupil A & 3.96 & 9.57  \\
\hline
\rule[-1ex]{0pt}{3.5ex}  Pupil B & 4.97 &  11.2  \\
\hline
\rule[-1ex]{0pt}{3.5ex}  Pupil C & 4.8 & 11.5 \\
\hline
\rule[-1ex]{0pt}{3.5ex}  Total A+B+C & 4.6 & 10.7  \\
\hline 
\end{tabular}
\end{center}
\end{table}

The respective SNR maps obtained with each screen are shown in Figure \ref{fig:snr_map_standard}. Also in these plots the gain in SNR with the new version of OLED is evident but it should be noted that there is a different distribution of signal on the pupils. 

\begin{figure}[H]
\begin{center}
\begin{tabular}{cc} 
   \includegraphics[width=8cm]{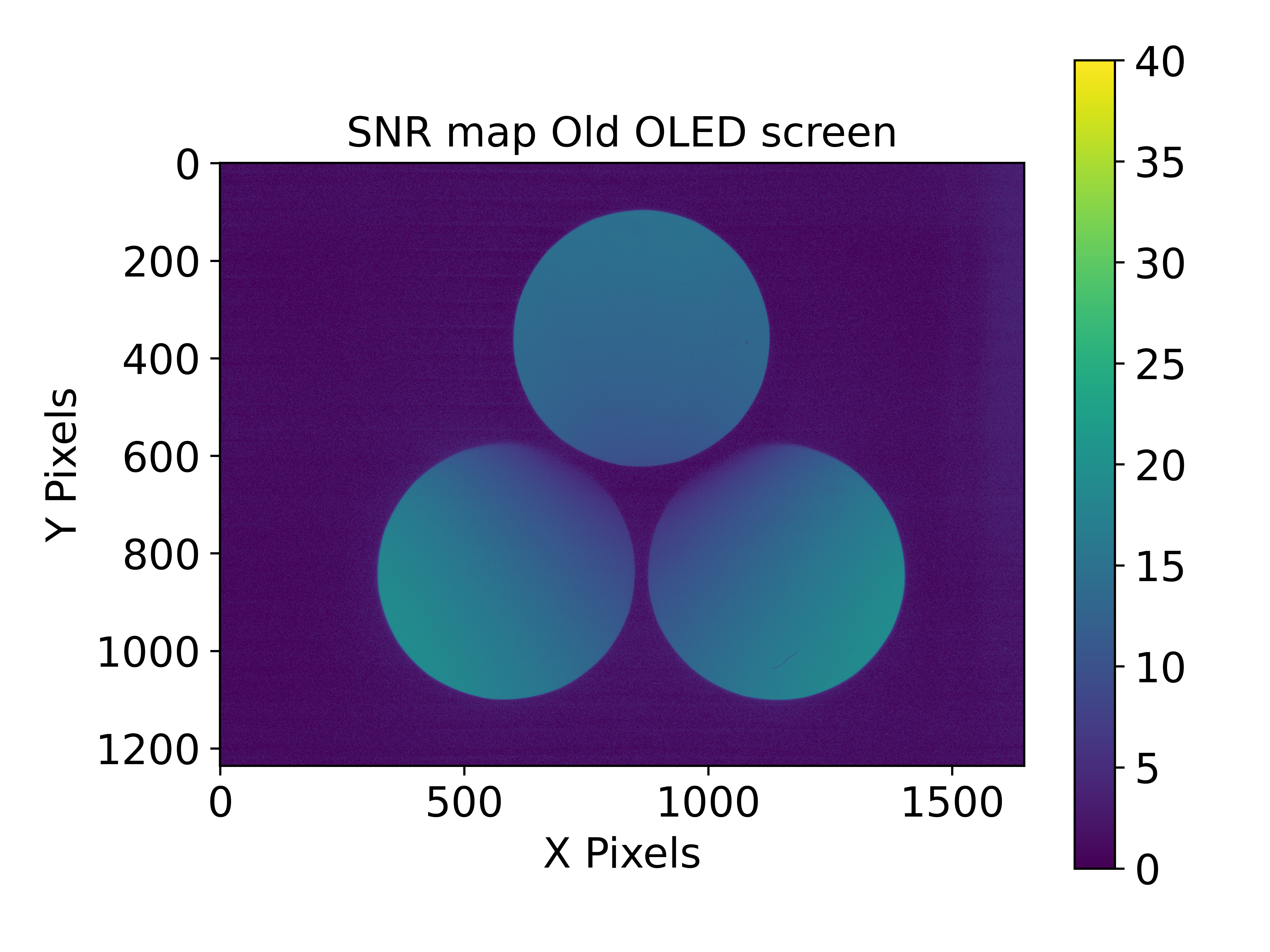} &
\includegraphics[width=8cm]{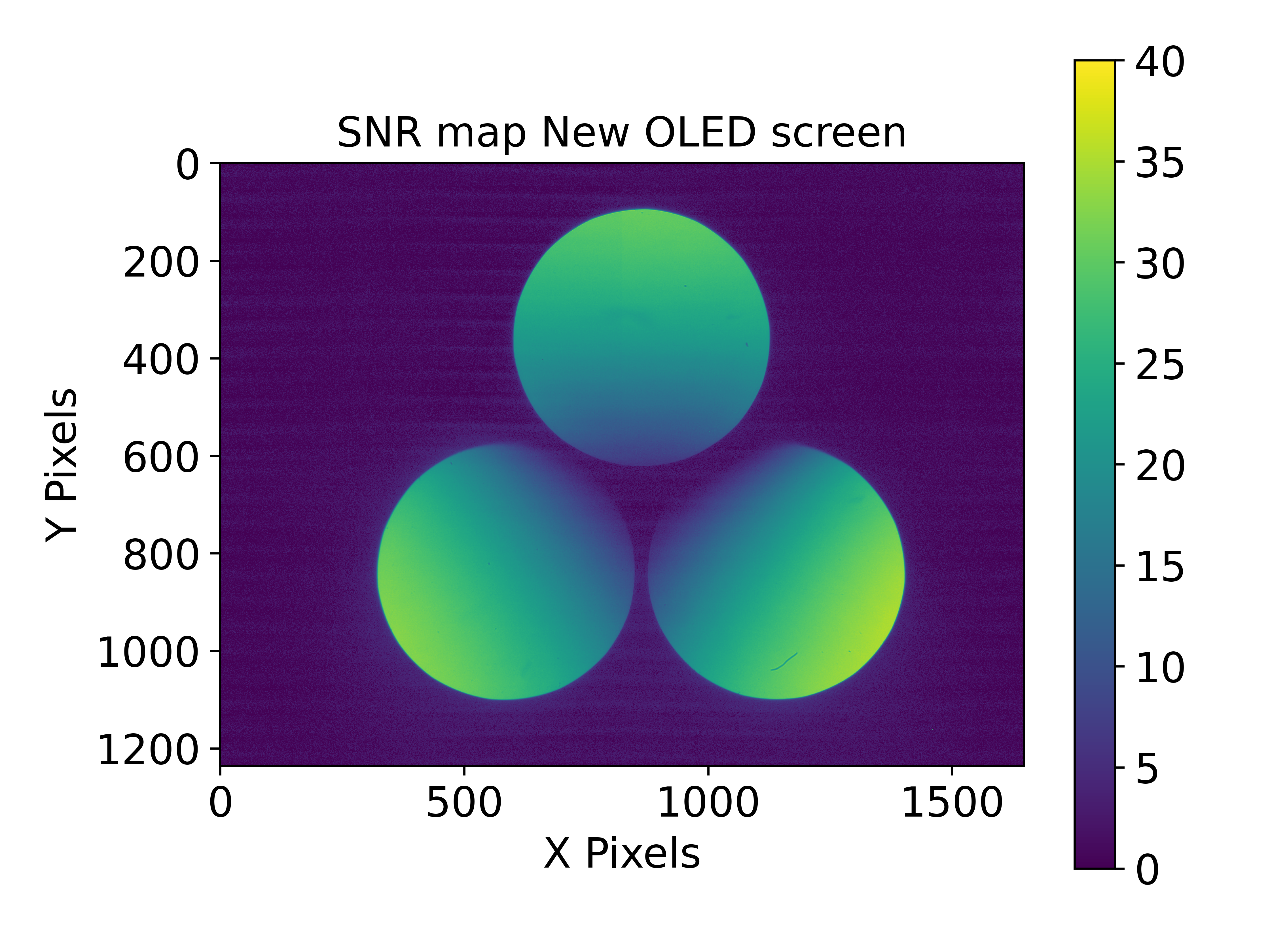}
   \end{tabular}

   \caption[example] 
   { \label{fig:snr_map_standard} (Left) SNR map of the image of the pupils of the I-WFS when it is aligned with the standard source on the old OLED screen. (Right) SNR map of the image of the pupils of the I-WFS when it is aligned with the standard source on the new OLED screen.}

\end{center}
\end{figure}

\vspace{-0.5cm}
\item \textbf{Study of the gradient along the pupils:}
To further investigate this different distribution we study the gradient of the distribution of light in the pupils in two different situations: only a transmitted beam arriving to the detector (the I-WFS is removed from the optical path) and all three pupils after reaching an aligned configuration of the standard source. For the case of having only a transmitted beam, the comparison between the gradients is plotted on Figure \ref{fig:just_transmitted_gradient}. A linear fit to the intensity is performed in order to better compare the trends. 

\begin{figure}[H]
\hspace{-0.3cm}
\begin{tabular}{c} 
   \includegraphics[width=1\linewidth]{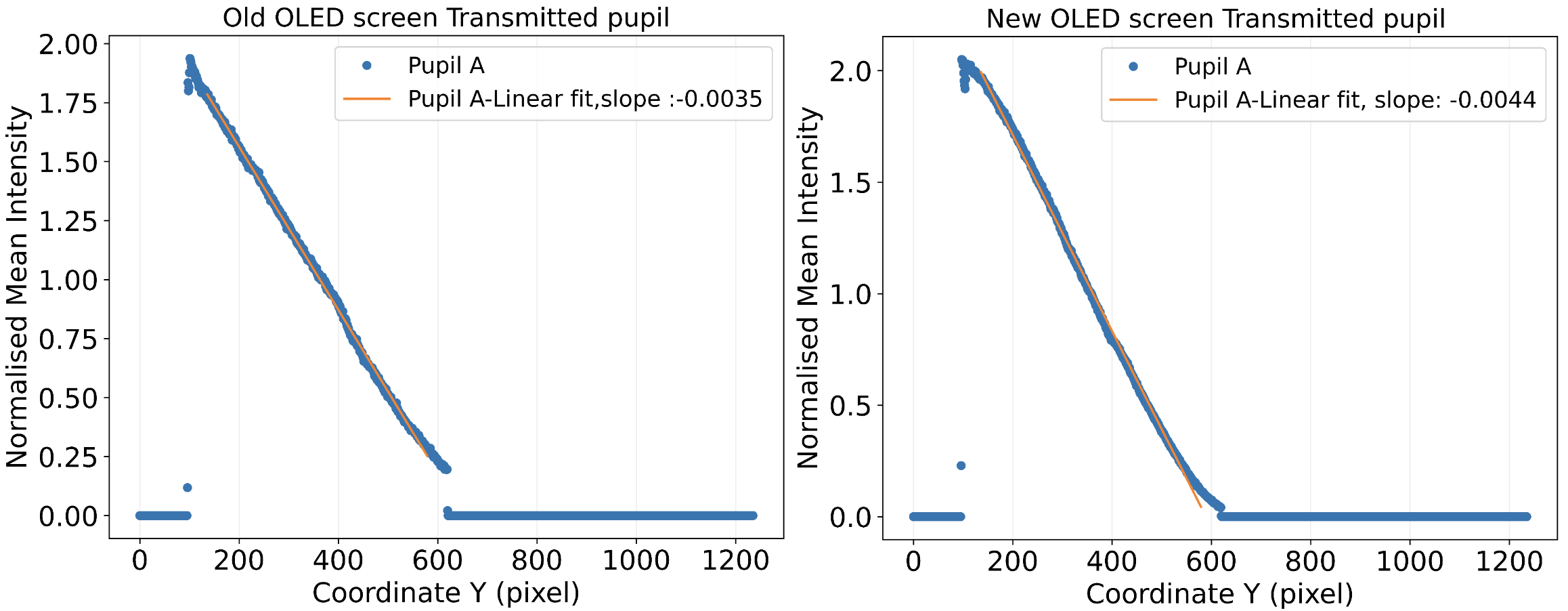} \\

   \end{tabular}

   \caption[example] 
   { \label{fig:just_transmitted_gradient} Gradient analysis of the transmitted pupil only. The average of the counts along each of the pupils is plotted, together with a linear fit to recover the slope. The x-axis contains the pixel coordinate along the pupil. The data is normalized with the average counts in that pupil. (Left) Transmitted pupil with the old screen. (Right) Transmitted pupil with the new screen. }

\end{figure}

\vspace{-0.5cm}
The gradient was decreased from -0.035 to -0.0044, being steeper in the new screen. The main difference between the two trends is at around 600 pixels, where the edge of the pupil is located. The old screen shows an increase of intensity from 0 to 0.25 between the background and the pupil's edge whereas this jump is nearly zero for the case of the new screen. The distribution of light inside the pupil is directly related with the angular (lateral/grazing) emission of the screens, which depends on its physical properties (coatings, screen thickness, material, etc). It is expected that different screens would have different emissions hence creating different gradients on the pupils.

For the case of having the three aligned pupils, the comparison between the gradients is plotted on Figure \ref{fig:three_aligned_gradient}. Besides the gradient for each pupil and respective liner fit, on these plots it is also plotted the gradient obtained from the sum of the fluxes on the three pupils.

\begin{figure}[H]
\hspace{-0.3cm}
\begin{tabular}{c} 
   
\includegraphics[width=\linewidth]{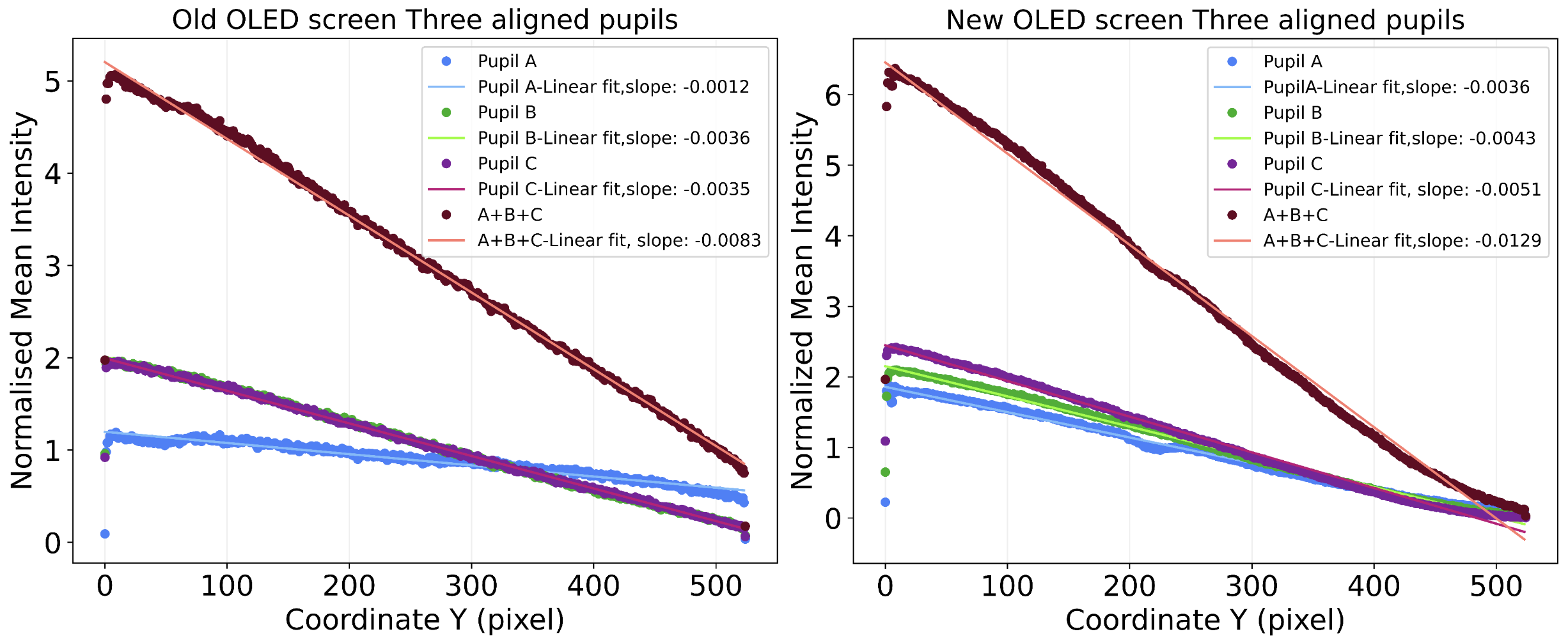}
   \end{tabular}

   \caption[example] 
   { \label{fig:three_aligned_gradient} Gradient analysis of the three aligned pupils. The average of the counts along each of the pupils is plotted, together with a linear fit to recover the slope. The x-axis contains the pixel coordinate along the pupil. For each pupil the data is normalized with the average counts in that pupil. (Left) Aligned pupils with the old screen. (Right) Aligned pupils with the new screen.}

\end{figure}

\vspace{-0.5cm}
A similar behaviour to the transmitted pupil is seen on the three aligned pupils: the gradients are steeper and there is a less visible jump between the background and the edge of the pupils. As a consequence, the methods of pupil detection could have more difficulties in finding the edge of the pupils.

\item \textbf{Study of the linearity of the alignment for different vertical shifts and calculation of the magnification of the system}: 
Given the approximate telecentric nature of the system, we anticipated that a movement of the source by one row (approximately 0.17 mm for the old screen and 0.21 mm for the new) would correspond to an analogous shift of the I-WFS along the Z axis. 

To test this behaviour and measure the magnification of the system, the following test was performed: we displayed sources with varying vertical shifts on both screens in random order and initiated the alignment process. Then, we evaluated the linearity of the alignment by fitting a linear function to the final Z values at the end of the alignment, as a function of the source vertical shift. The data fit is depicted in Figure \ref{fig:vs_z}. 

The fit parameter that is related to the magnification of the system was determined to be a = -0.174 mm and a=-0.212 mm. Notably, these values align with expectations, indicating that each increment of VS by 1 corresponds to a Z shift of -0.174 mm and -0.212 mm, effectively the pixel pitch. The standard deviation of this dataset from the linear fit stands at $\sigma$ = 0.039 mm and $\sigma$ = 0.007 mm. We believe that the larger $\sigma$ on the old OLED screen is due to the wearing off of the most used rows of the screen, which instead is not present on the newly acquired new screen. 

The magnification of the system is then obtainable by dividing these values by the corresponding pixel pitch. We obtain from the data of the old screen a magnification of 1.02 and 1.01 from the data of the new screen, which is according to our expectations a confirmation that the optical layout of the bench is 1:1.

\begin{figure}[H]
\hspace{-0.3cm}
\begin{tabular}{c} 
   \includegraphics[width=\linewidth]{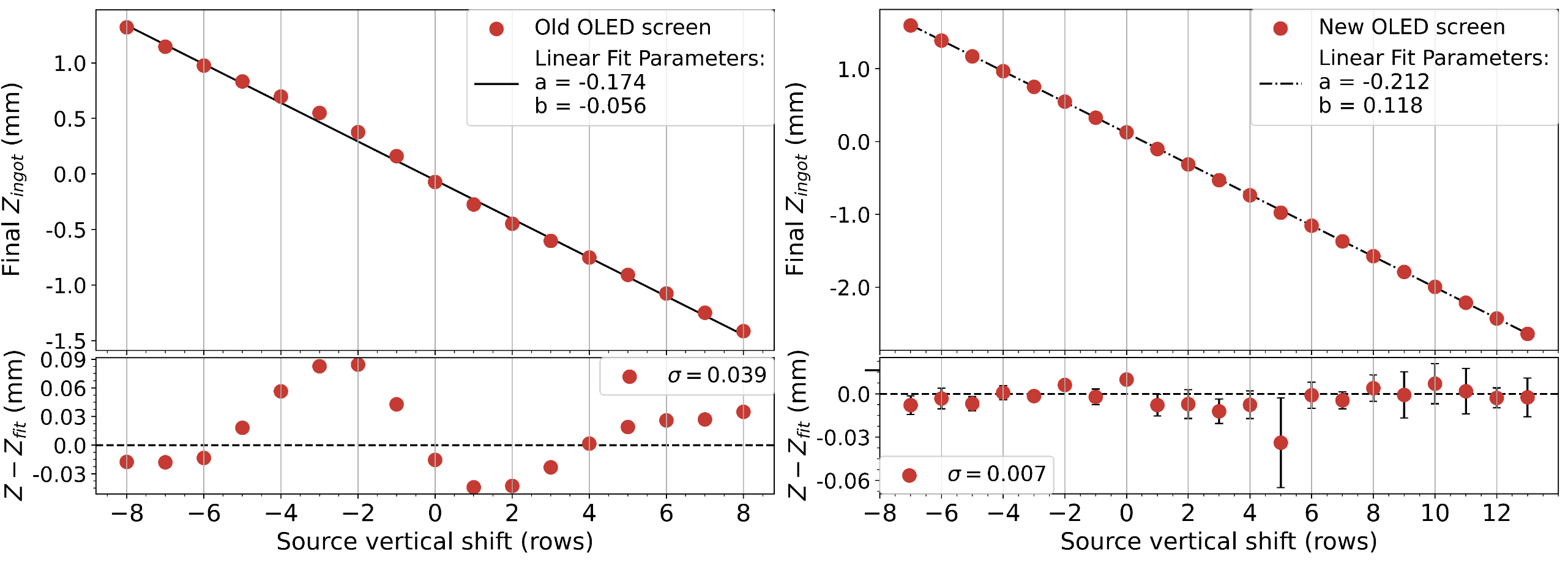}
   \end{tabular}

   \caption[example] 
   { \label{fig:vs_z} Results from alignments of sources with different vertical shifts on the two screens. The x-axis displays the vertical shift of the source, in number of rows on the screen. The y-axis represents the final Z axis position of the I-WFS, in mm. (Left) Old screen. Top: Linear fits performed to the data, with best fit parameters $a = -0.174$ mm/row and $b = -0.056$ mm (moving screen) Bottom: The residuals of the fit have a standard deviation of $\sigma= 0.039$ mm. (Right) New screen. Top: Linear fits performed to the data, with best fit parameters $a = -0.212$ mm/row and $b = -0.118$ mm (moving screen) Bottom: The residuals of the fit have a standard deviation of $\sigma= 0.007$ mm.
 }

\end{figure}

\end{enumerate}

\vspace{-0.5cm}

\subsection{Sensitivity of the I-WFS to brightness variations}

\begin{figure}[H]
\begin{center}
\begin{tabular}{cc} 
   \includegraphics[width=7cm]{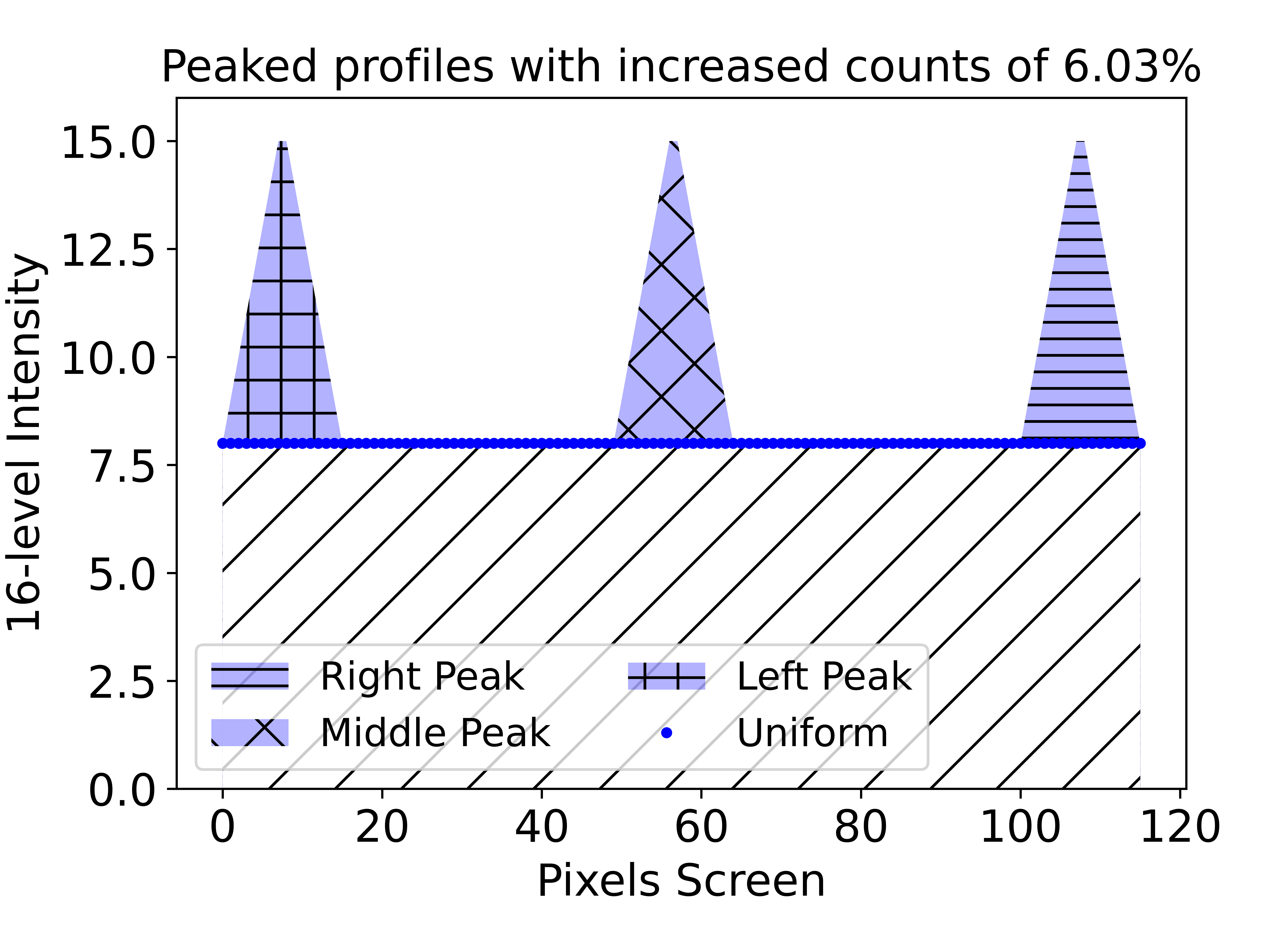} &
\includegraphics[width=7cm]{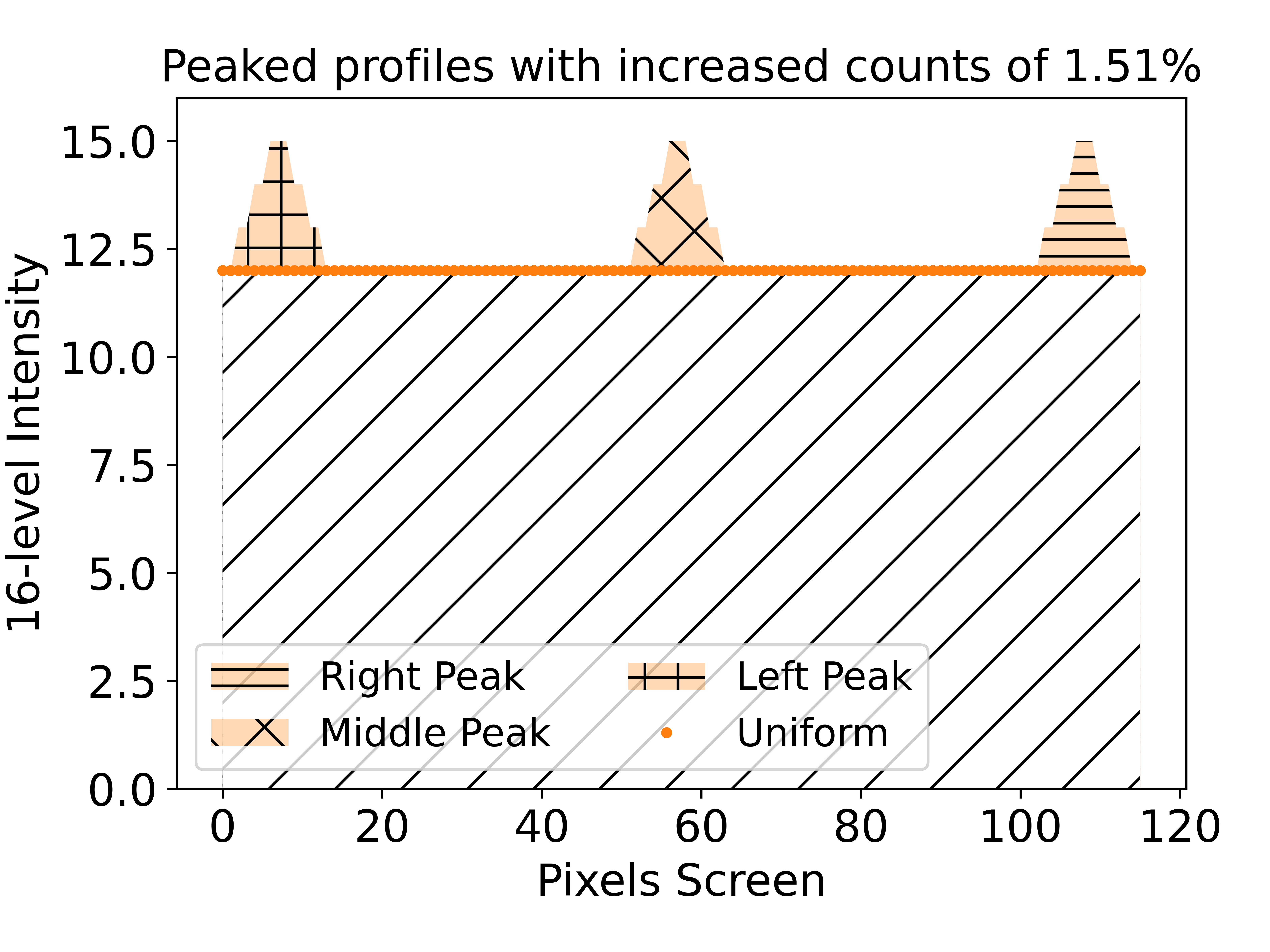} \\
   \includegraphics[width=7cm]{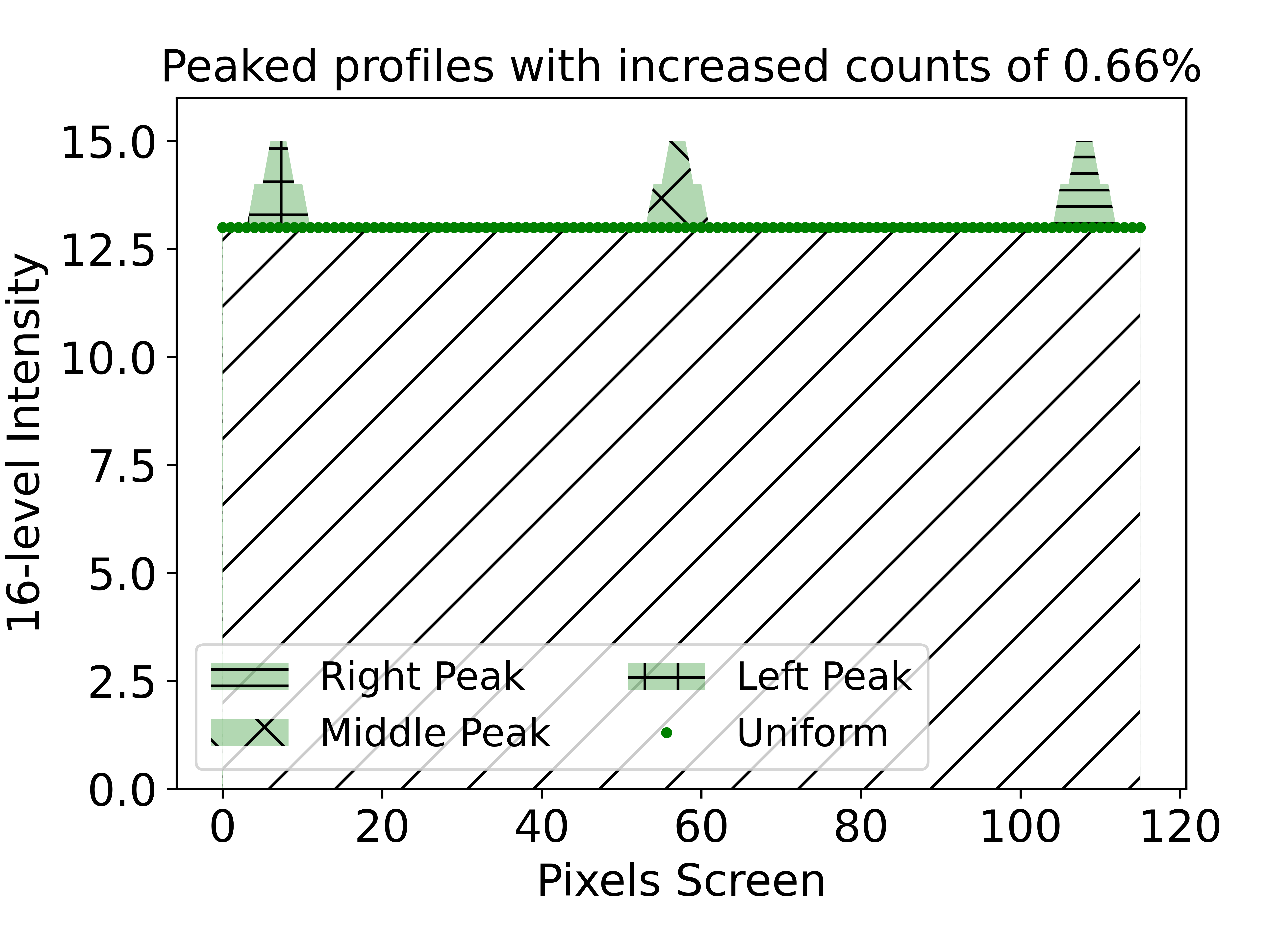} &
\includegraphics[width=7cm]{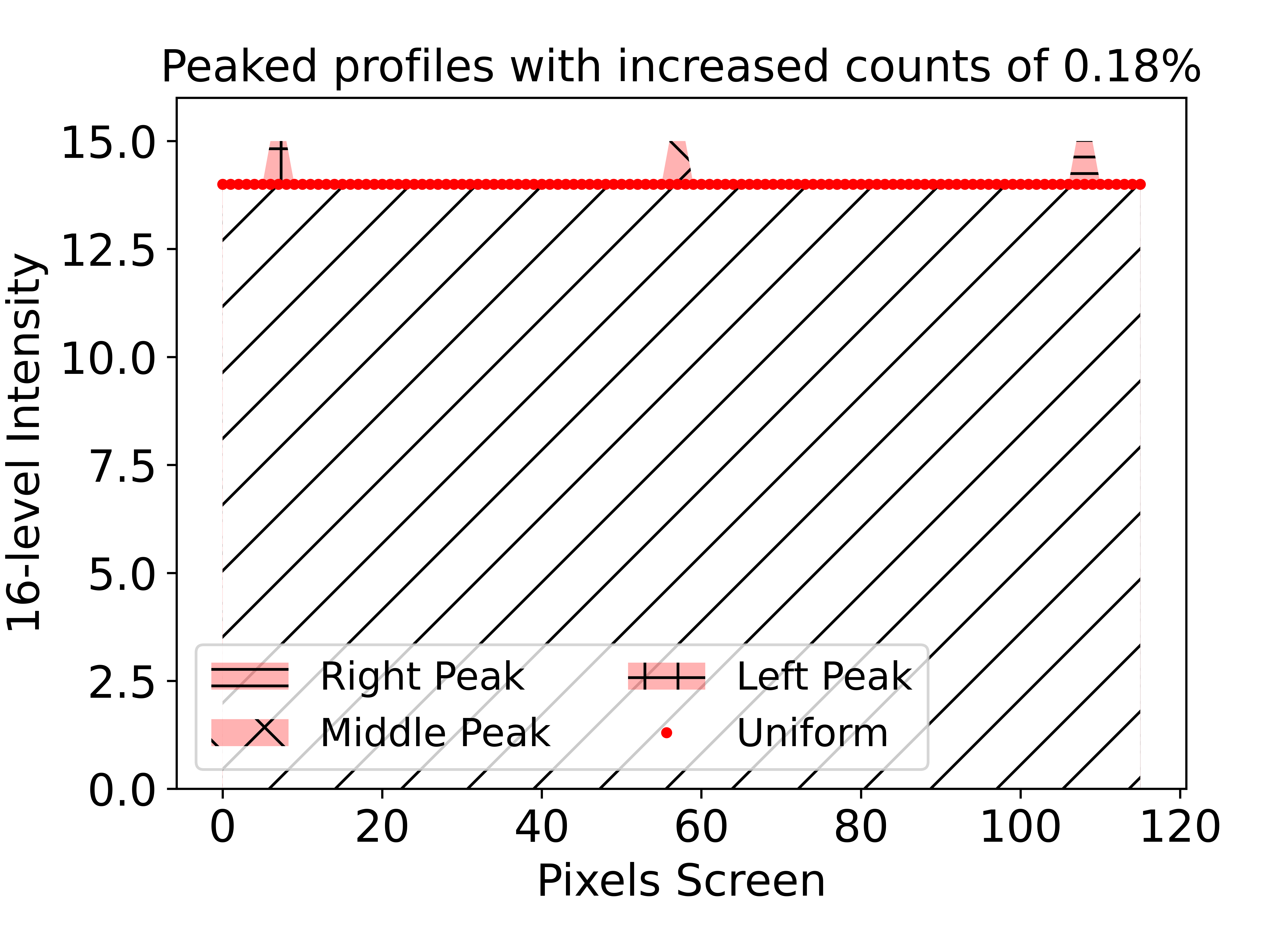}  \\
   \end{tabular}

   \caption[example] 
   { \label{fig:peaks_dheeraj} Uniform profiles of four different intensities on a scale of 0 to 15. On top, three profiles area added each with a gaussian peak of maximum intensity 15. (Top Left) Intensity 8, which means the increase in area/intensity due to each gaussian is $6.03$\% and is here highlighted by the blue shaded area. (Top Right) Intensity 12, which means the increase in area/intensity due to each gaussian is $1.51$\% and is here highlighted by the orange shaded area. (Bottom Left) Intensity 13, which means the increase in area/intensity due to each gaussian is $0.66$\% and is here highlighted by the green shaded area. (Bottom Right) Intensity 8, which means the increase in area/intensity due to each gaussian is $0.18$\% and is here highlighted by the red shaded area. }

\end{center}
\end{figure}

With the new capabilities of this experimental set-up, we proceeded to study ability of the I-WFS to align itself when the source is not uniform but has a differential brightness profile. To study the sensitivity, we investigated what is the minimum increase in brightness of the LGS that will cause a change in the alignment position of the I-WFS. Moreover, we question whether the sensitivity is the same if there is an increase of brightness anywhere along the source (far end, middle end, closer end). To understand this question, the alignment of the I-WFS to a uniform source of 4 levels of brightness - 8, 12, 13 and 14 is done and compared to the aligned position of the same profile with a peak of brightness of 15, positioned on three positions along the length of the source. The profiles with the gaussian peak positioned at the three positions are displayed in Figure~\ref{fig:peaks_dheeraj}.

When comparing the final aligned position of each of the above mentioned profiles with the aligned position of the respective uniform profile, we can conclude that the defocus of the system (Y coordinate) is affected when the relative increase of counts in the profile is at least 0.66\% and that there is not a position along the source where the system is more sensitive to (the Y shift is symmetric with left and right peaks). The X coordinate is affected mostly when the peak is located in the right end (closer to the pupil) and the shifts in Z coordinate are relatively small and can be neglected.

\vspace{-0.5cm}

\begin{figure}[H]
\hspace{-0.3cm}
\begin{tabular}{c} 
   \includegraphics[width=\linewidth]{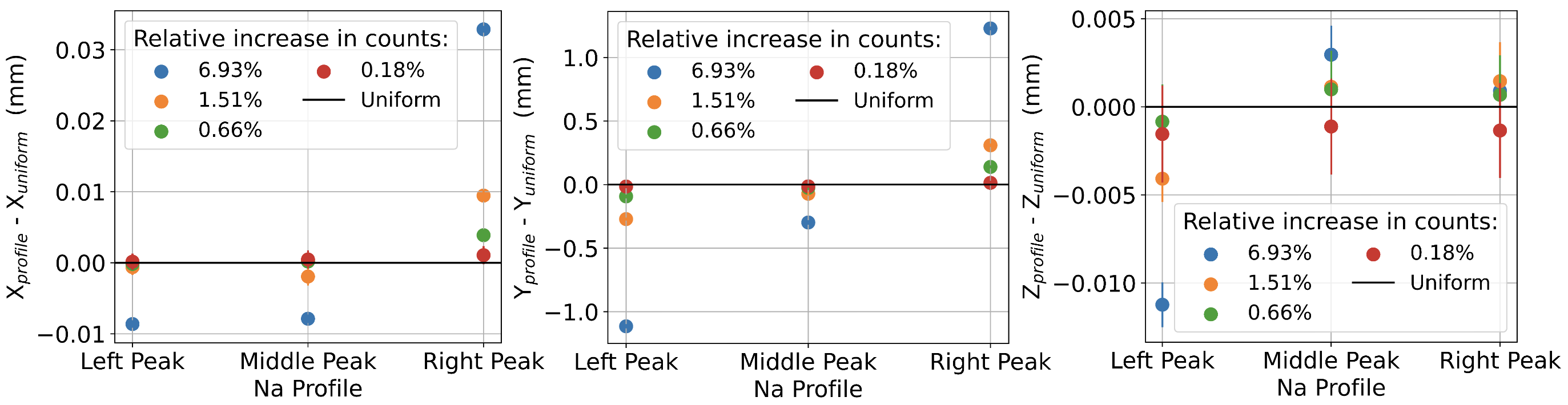} 
 \end{tabular}

   \caption[example] 
   { \label{fig:example} Comparison of the final aligned position of each of the above mentioned profiles with the aligned position of the respective uniform profile,$X_{profile}-X_{uniform}$. (Left) X axis. (Middle) Y axis. (Right) Z axis. }

\end{figure}

\vspace{-0.5cm}

\begin{figure}[H]
\begin{center}
\begin{tabular}{cc} 
   \includegraphics[width=7cm]{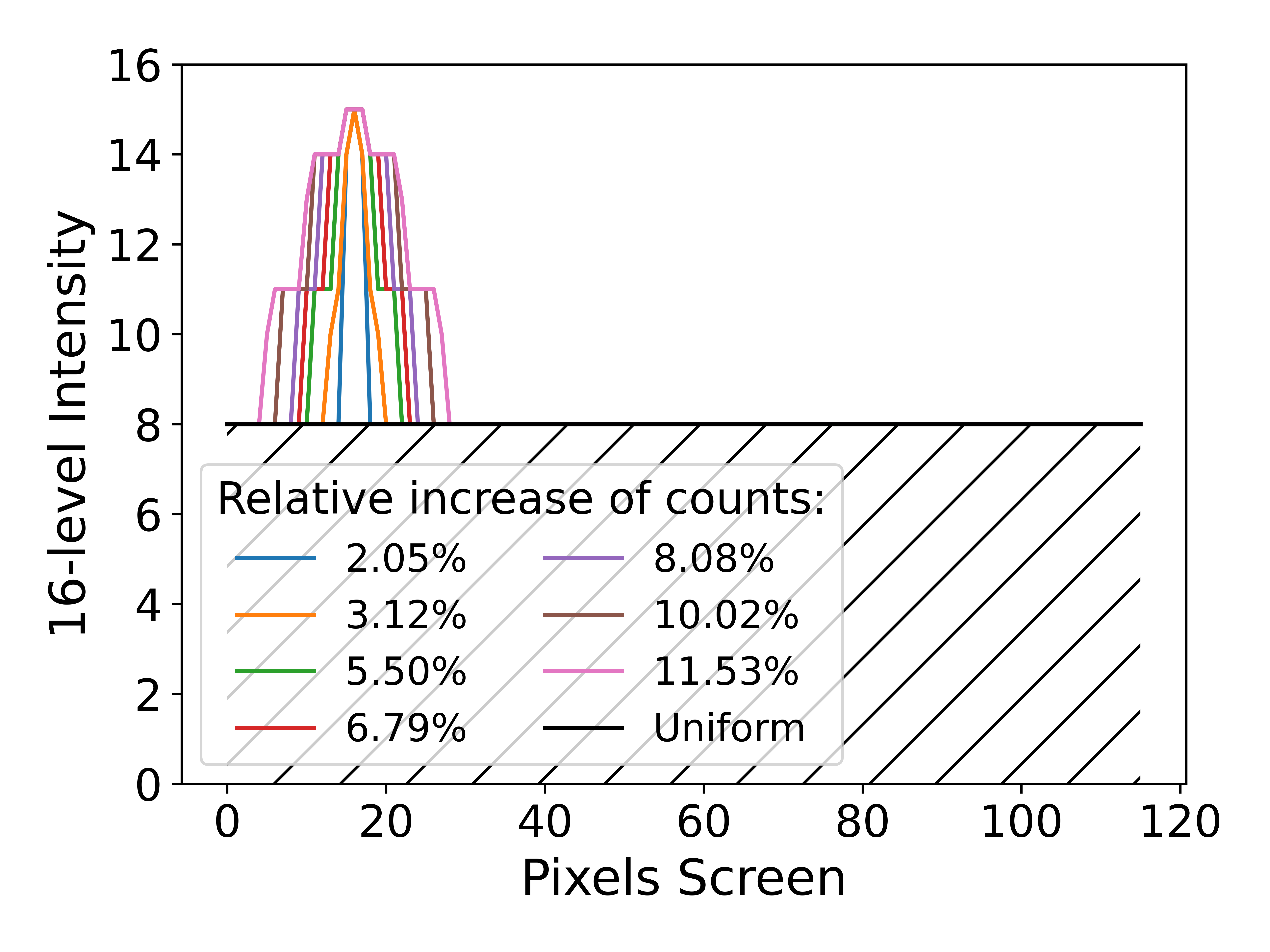} &
\includegraphics[width=7cm]{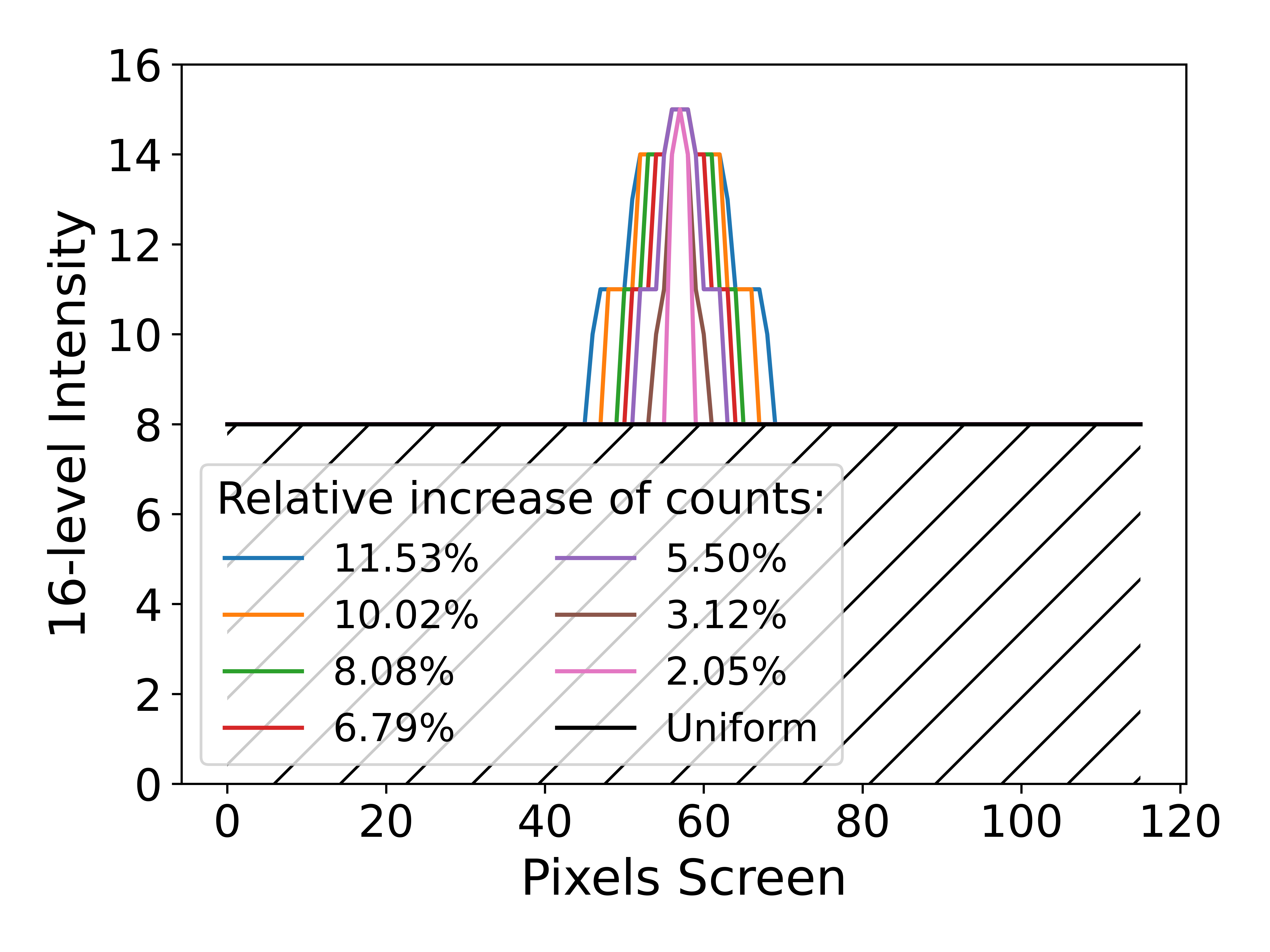}  \\
   \multicolumn{2}{c}{ \includegraphics[width=7cm]{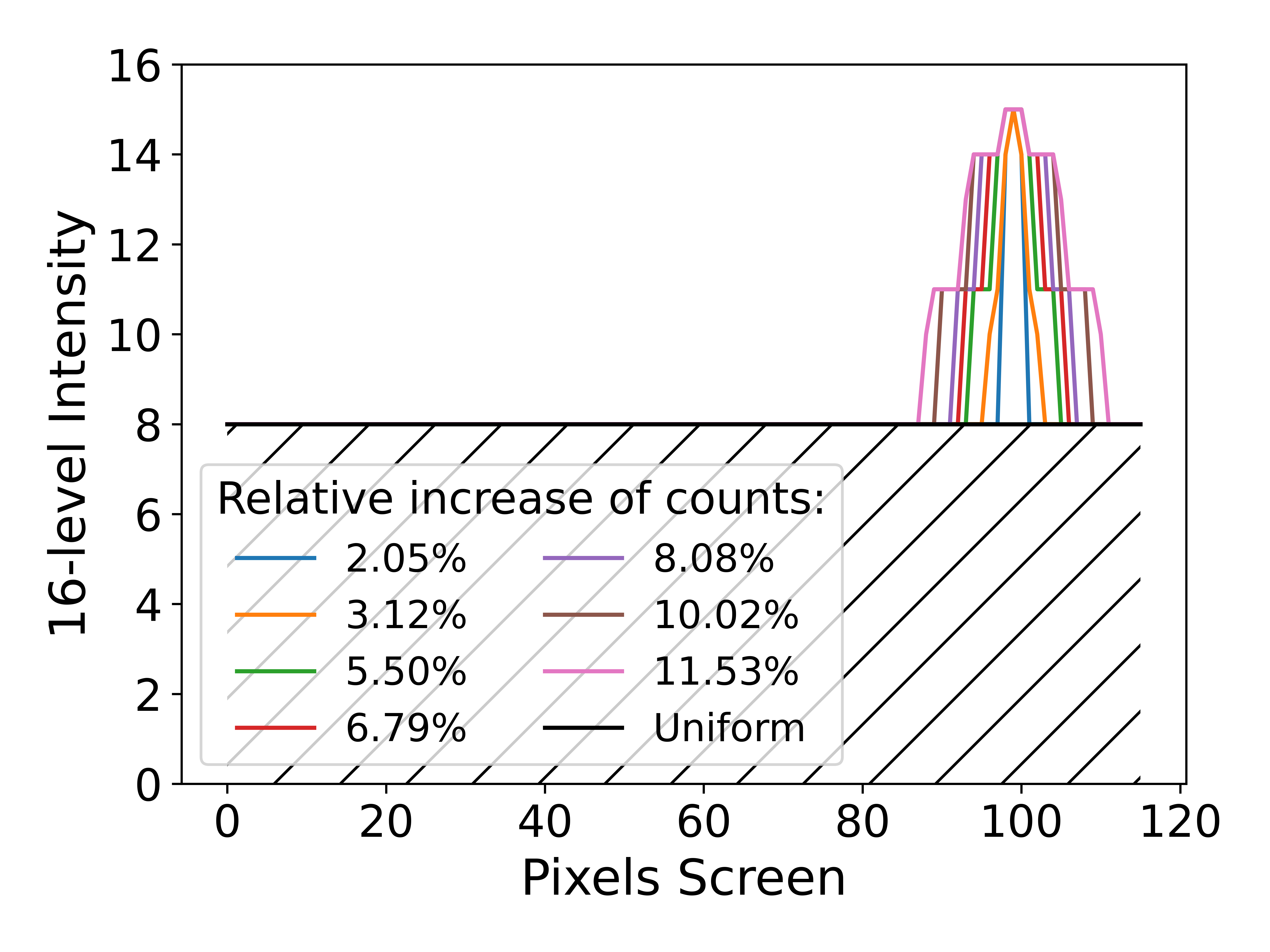}} \\
   
\end{tabular}

   \caption[example] 
   { \label{fig:peaks_width} Uniform profile of intensity 8 on a scale of 0 to 15. On top, seven profiles are added, each with a gaussian peak of maximum intensity 15 and different width. (Top Left) Peaks added in the far end of the LGS. (Top Right) Peaks added in the middle of the LGS. (Bottom) Peaks added in the close end of the LGS.}

\end{center}
\end{figure}

Another question we aimed at understanding concerned the width of the intensity peaks applied to the screen, that correspond to layers in the sodium layer where there is a peak of sodium concentration. We were interested in measuring how thick/thin these have to be in order for the I-WFS to sense a difference in the aligned position. To test this, we again create artificial profiles over a uniform profile, that have a gaussian peak. This time, however, we change the width of the gaussian instead of the level of the uniform source. We again apply these profiles on both far end, middle and close end to test if the system is equally sensitive along the LGS extension (Figure \ref{fig:peaks_width}). The main conclusion of this analysis is that the main factor that affects the alignment is the relative increase in counts and not the width of the gaussian peak.

\vspace{-0.2cm}

\vspace{-0.5cm}

\begin{figure}[H]
\begin{center}
\begin{tabular}{c} 
   \includegraphics[width=10cm]{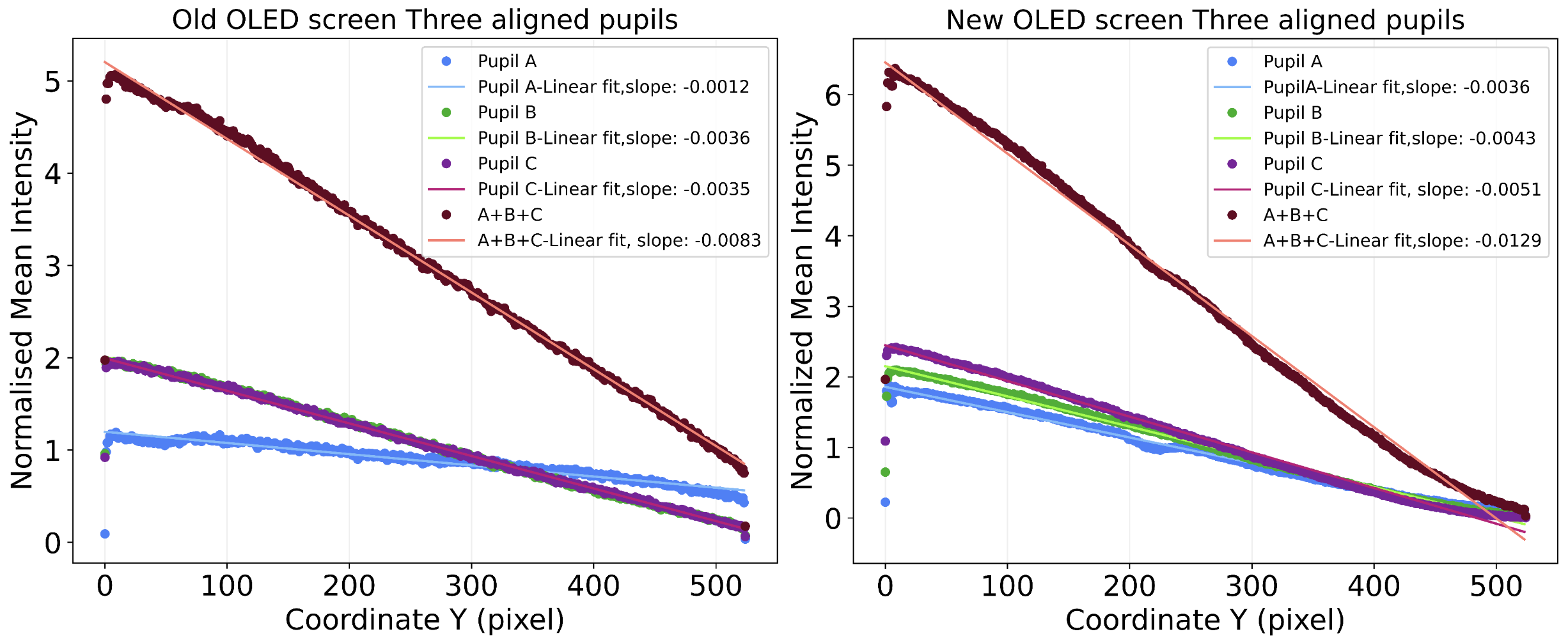} \\
\includegraphics[width=13cm]{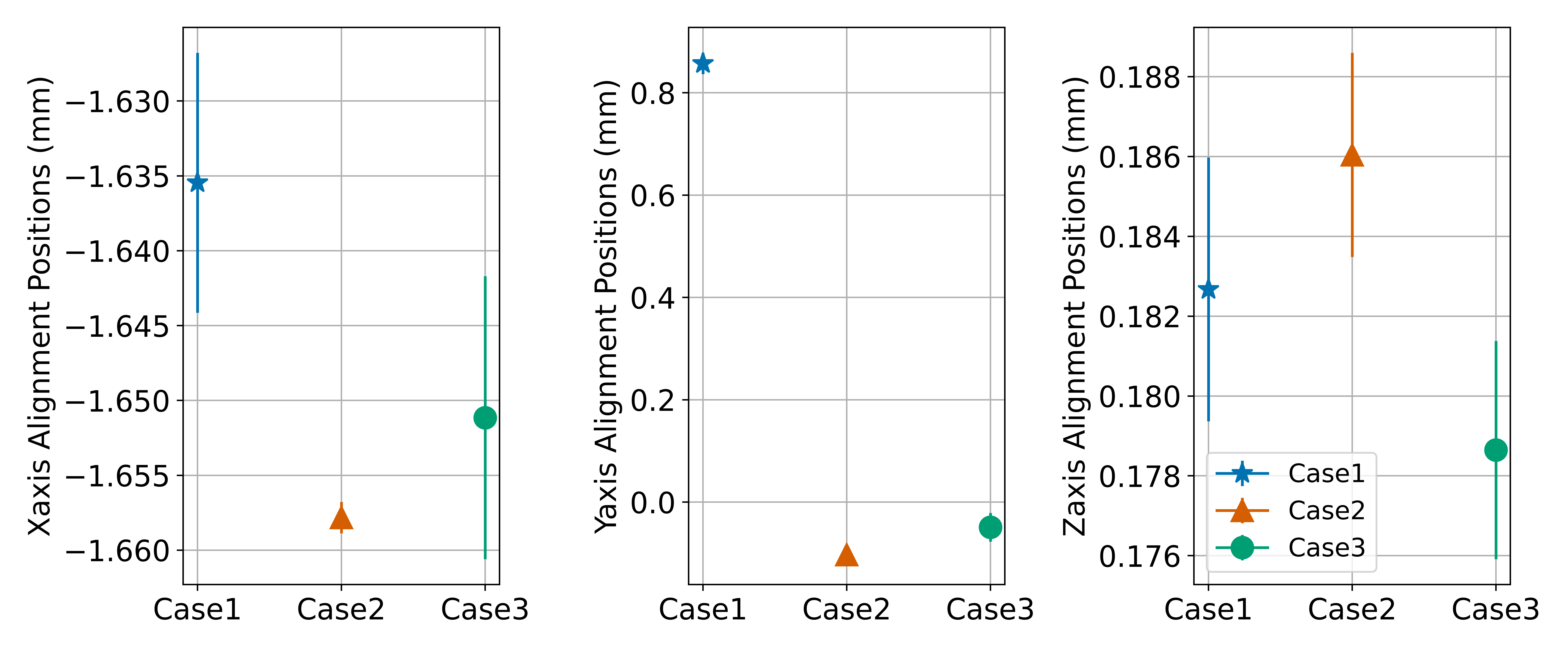} 
   \end{tabular}

   \caption[example] 
   {\label{fig:all_cases} (Top) The three cases with sodium profiles obtained from the sodium layer data, previously shown on Figure \ref{fig:all_cases}. They correspond to measurements taken in three intervals of 2.5 hours  on the same night. (Bottom) Average aligned positions in the X, Y and Z axis of the I-WFS hexapod for the three profiles.}

\end{center}
\end{figure}

Having understood the sensitivity of the I-WFS alignment, we moved on to analysing how this alignment is affected using real profiles taken throughout one night. The nightly variation of the profiles is represented by the three cases displayed on Figure \ref{fig:all_cases}, which are separated by 2.5 hours. In the same Figure, we show how the aligned position of the I-WFS is altered for each of the different cases. The biggest effect is a change in defocus of the order of 0.8 mm.

\begin{figure}[H]
\begin{center}
\begin{tabular}{c} 
   
\includegraphics[width=0.8\linewidth]{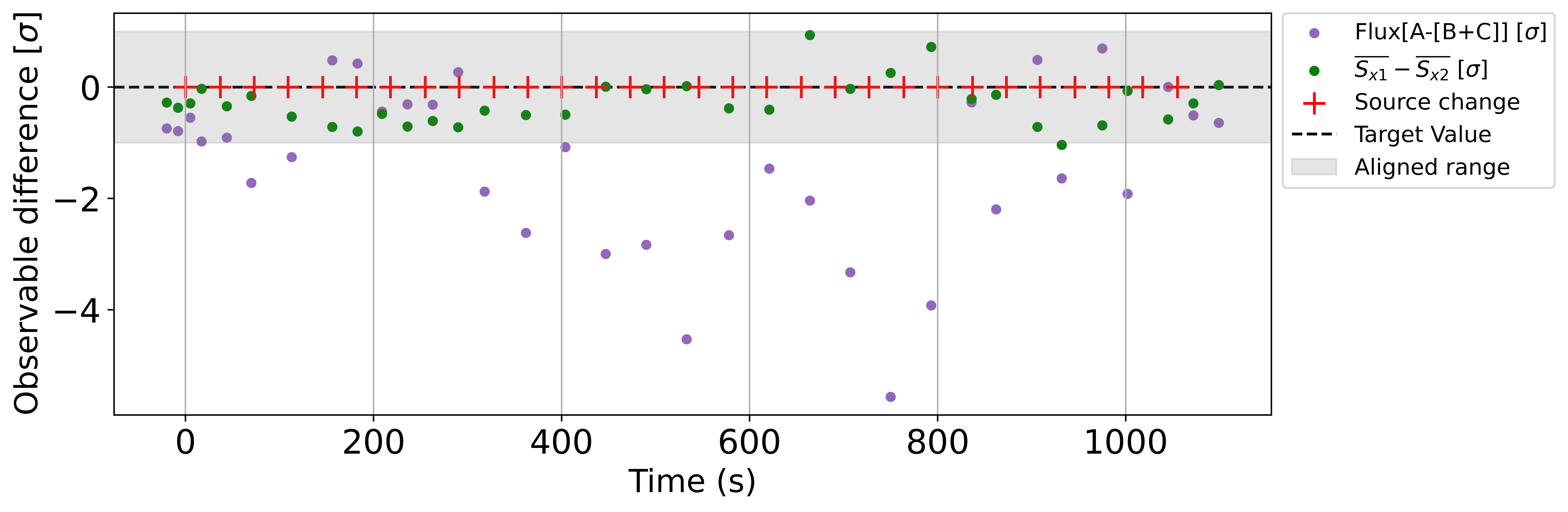} 
   \end{tabular}

   \caption[example] 
   { Evolution of Flux$(A-B-C)$  and $\overline{S_{x1}}  - \overline{S_{x2}} $ observables as the source is changing sequentially on the screen. The red crosses indicate when the source is changed and the green and purple dots the measurement of the observables of the pupil.  \label{fig:source_change_30s}  }

\end{center}
\end{figure}

This implies that, in the course of one night, the system needs to readjust its alignment. Next, looking at faster profile changes, we investigated whether the alignment of the system is affected by the 5-minute changes of the profile. To investigate this, we first align the I-WFS to the first set of data of the night, and then, without moving the system, we simply change the profile on the screen every 30 seconds. For every source change, we measured the observables of the system and calculate how far from the tolerance of alignment ($\sigma$) we are with the current source on the screen. The plot in Figure \ref{fig:source_change_30s} aims to show the evolution of two observables as the source is changing sequentially on the screen. We noticed how these two observables, Flux$(A-B-C)$  and $\overline{S_{x1}}  - \overline{S_{x2}} $, are the mostly affected by the change in source, even though the latter does not oscillate more than $1\sigma$. The first one, oscillates up to $5\sigma$, which again is informing us that the I-WFS needs to be continuously adjusted at least in defocus. 

\vspace{-0.5cm}

\vspace{-0.7cm}

\subsection{Introducing low-order aberrations}
A Deformable Lens (DL) positioned in the pupil plane was previously tested in Ref.~\citenum{DiFilippo2022}. A collection of tests was performed with the DL and can be reviewed on the works of Ref.~\citenum{2020SPIEKalyan}. The DL is the AOL1825 developed by Dynamic Optics \cite{Mal_DO}, with a clear aperture of 25.5mm and 18 actuators (9 on each side of the lens). This device, also referred to as objective integrated multi-actuator adaptive lens, is able to introduce low-order known aberration terms to the wavefront, and presents two notable advantages compared to other possible solutions: it does not require the modification of the existent optical path and its installation is extremely straightforward, common off-the-shelf lab parts can be used to hold it. The DL is able to apply the first 18 Zernike modes and can be controlled both using its own software or the Matlab or Python custom libraries developed by the owner. 

When inserted on the optical path, the DL should act as a window for the incoming light, if no aberrations are introduced. Using a 4D PhaseCam 4030 Interferometer we are currently calibrating this lens, analysing its stability and characterising its flat configuration. We plan to incorporate it temporarily on the optical path and study the effect of its presence on the alignment of the I-WFS.

\section{CONCLUSIONS AND FUTURE WORK}
We presented here the main developments on the laboratory tests of the I-WFS. For starters, a search algorithm was developed to allow the I-WFS to recover from situations where no pupil, only one or only two pupils are illuminated. Then, the alignment procedure has been refined to correct for a Y axis overshoot and reach a more stable aligned position. 

In addition, the test bench has been updated to allow simulations of LGS that include sodium profile variability along the LGS length, using OLED technology. A comparison between two OLED elements (the old screen and the new screen) was performed to understand the capabilities of a newly acquired OLED screen. This element maintains a low background emission from dark pixels while increasing the average SNR of the aligned pupils two fold. After studying the sensitivity of the I-WFS to the presence of a non uniform profile along the LGS extension, we successfully aligned the I-WFS with profiles originating from real data. We show that the alignment is stable and requires mostly an adjustment in defocus as the profile evolves throughout one night. 

Finally, low-order aberrations are expected to be introduced in the near future and a full study of the sensitivity and linearity of the I-WFS will follow. On-sky testing is currently being planned, and to reach that phase of this ambitious project we hope to introduce a deformable mirror on the test bench and perform closed-loop operations.

\vspace{-0.5cm}

\acknowledgments 
 
We acknowledge the ADONI Laboratory for the support in the development of this project. Funding to support the lab activities described in this paper came also from the INAF Progetto Premiale "\textit{Ottica Adattiva Made in Italy per i grandi telescopi del futuro}”. We also show great appreciation to Angel Otarola, from TMT International Observatory, for making the sodium profile data available to us. We thank Tommaso Furieri and Stefano Bonora for the support provided with the multi-actuator adaptive lens. Finally, we extended our appreciation to Claudio Pernechele for the help provided by giving us the chance of using the \textit{Ocean Insight} spectrometer in our tests.

\bibliography{report} 
\bibliographystyle{spiebib} 

\end{document}